\documentclass[12pt, letterpaper]{article}

\usepackage[margin=1in]{geometry}%
\usepackage{setspace}%
\usepackage{lineno}%
\usepackage{graphicx}%
\usepackage{rotating}
\usepackage{multirow}%
\usepackage{amsmath,amssymb,amsfonts}%
\usepackage{amsthm}%
\usepackage{mathrsfs}%
\usepackage[title]{appendix}%
\usepackage{xcolor}%
\usepackage{algorithm}%
\usepackage{algorithmicx}%
\usepackage{algpseudocode}%
\usepackage{listings}%
\usepackage{tikz}
\usepackage{subcaption}
\usepackage[round]{natbib}
\usepackage{authblk}
\usepackage{url}
\usepackage{colortbl}

\begin{document}

\title{Predicting fine-scale taxonomic variation in landscape vegetation using large satellite imagery data sets}

\author[1]{Henry Scharf\thanks{corresponding author: henryrscharf@gmail.com}}
\author[1]{Jonathan Schierbaum}
\author[2]{Hana Matsumoto}
\author[2]{Tim Assal}

\affil[1]{Department of Mathematics and Statistics\\
  San Diego State University\\
  5500 Campanile Drive\\
  San Diego, CA 92812}

\affil[2]{Department of Geography\\
  Kent State University\\
  800 E. Summit Street\\
  Kent, OH 44242}

\maketitle

\begin{abstract}
Accurate information on the distribution of vegetation species is used as a proxy for the health of an ecosystem, a currency of international environmental treaties, and a necessary planning tool for forest preservation and rehabilitation, to name just a few of its applications. However, direct, extensive observation of vegetation across large geographic regions can be very expensive. The extensive coverage and high temporal resolution of remote sensing data collected by satellites like the European Space Agency's Sentinel-2 system could be a critical component of a solution to this problem. 

We propose a hierarchical model for predicting vegetation cover that incorporates high resolution satellite imagery, landscape characteristics such as elevation and slope, and direct observation of vegetation cover. Besides providing model-based predictions of vegetation cover with accompanying uncertainty quantification, our proposed model offers inference about the effects of landscape characteristics on vegetation type. Implementation of the model is computationally challenging due to the volume and spatial extent of data involved. Thus, we propose an efficient, approximate method for model fitting that is able to make use of all available observations. We demonstrate our approach with an application to the distribution of three post-fire resprouting deciduous species in the Jemez Mountains of New Mexico.
\end{abstract}

\textbf{keywords:} species distribution model, remote sensing, Jemez Mountains, Bayesian, spatial statistics, vegetation disturbance

\section{Introduction}\label{sec:introduction}

\subsection{Overview}

Land cover and vegetation information derived from remote sensing data facilitates a wide variety of geographical analyses and science applications. These data have been used to study the biodiversity of forests as a proxy for ecosystem health \citep{AREK2017}, monitor the progression of wildfires \citep{CROW2019} and invasive diseases \citep{bradley2014remote, HAUG2016}, and study the long-term effects of climate change \citep{WYLI2014}, while also playing a critical role in local conservation planning and international forest protection agreements \citep{PATE2005}. Even in the current era of high-resolution satellite imaging and sophisticated image processing techniques, there remain heterogeneous ecosystems that are too varied across space to be characterized to the degree of specificity desired by all stakeholders. For example, the well known National Land Cover Database maintained by the United States Geological Service uses a classification system with nine broad land cover categories (e.g., water, forested upland) and 21 sub-categories based on shared structural, physiological, and phenological traits (e.g., deciduous forest, evergreen forest, mixed forest). However, information about the distribution of particular species within a forest landscape, information that is critical for monitoring ecosystem health and implementing sustainable forest management practices, remains elusive at efficient scales \citep{ASSA2015}.

The current era of satellite imaging has become synonymous with high temporal resolution as well. As an example, the twin polar-orbiting satellites of the Sentinel-2 mission under the European Space Agency's (ESA) Copernicus program have a revisit period ranging from five days at the equator to a single day at higher latitudes. Researchers have access to vast repositories of historical imaging data via geospatial information aggregators like Google Earth Engine, with the current day's imaging data available in less than 24 hours. This stream of freely available satellite data provides opportunity to create rich data sets that better express the phenological differences of vegetation by incorporating many remote sensing images at once, thereby making it possible to map more thematically detailed classes with higher accuracy.  

Remote sensing imagery provides information about vegetation cover through multiple pathways. The proportion of solar radiation reflected off of vegetation cover varies across seasons and frequencies of light in patterns that are distinct even among closely related species. Landscape characteristics such as elevation and slope, derived from digital elevation models like those provided by the U.S. Geological Survey \citep{USGS_map}, can also help explain observed patterns of occurrence to better understand which locations in a region are most associated with specific vegetation types.

We develop and apply a Bayesian hierarchical species distribution model \citep[SDM; e.g.,][]{guisan_predicting_2005, GELF2006} with the goal of inferring vegetation cover type at a fine taxonomic resolution in the fire-scarred landscapes of Jemez Mountains in the Northwest quadrant of New Mexico. SDMs have been used extensively to analyze species distributions across landscapes and offer a pathway for the classification of remote sensing data by species \citep{ZIMM2007, he2015will}. Our proposed model allows for the incorporation of remotely sensed observations of both reflected solar radiation and landscape features such as elevation, slope, and heat loading, and yields information about habitat suitability for each vegetation type as a function of these inputs. We evaluate the consistency and practical value of our proposed model in a simulation study, and apply it to the study of three post-fire resprouting deciduous species (quaking aspen, New Mexican locust, and a mix of Gambel and wavy leaf oak species) over tens of thousands of sites. Implementing our model for the motivating application requires a novel approach to accommodate the vast number of satellite imagery measurements gathered via remote sensing.

\subsection{Study area}
\begin{figure}[ht]
  \includegraphics[width=\linewidth]{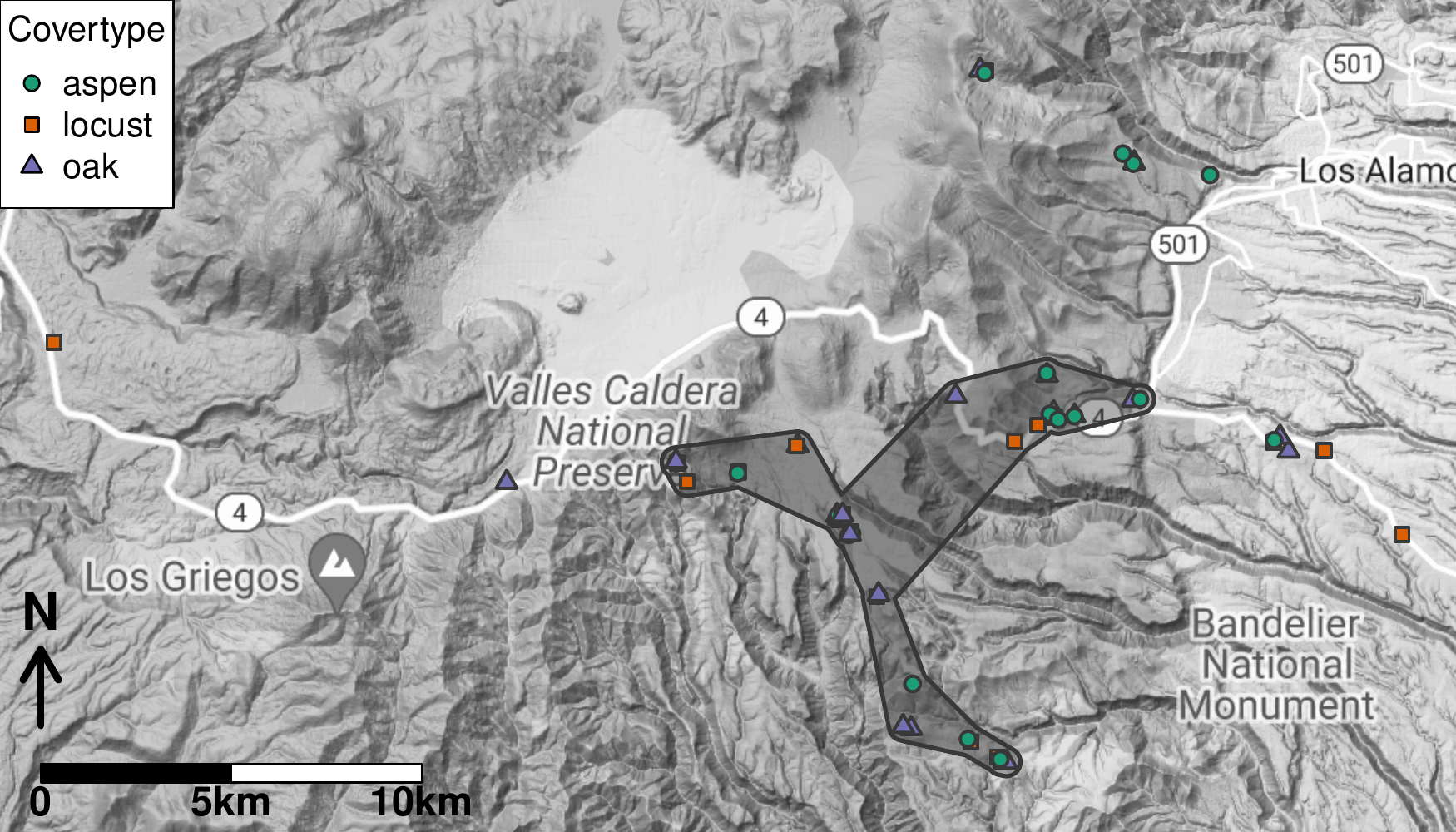}
  \caption{Study area. Points represent locations where georeferenced photos were used to identify vegetation cover type, with color/shape corresponding to type. The ``Y''-shaped polygon represents the region of interest for cover type predictions.}
\label{fig:study_area}
\end{figure}

The study area for our motivating application is located within the Jemez Mountain Range in the Northwest quadrant of New Mexico, approximately 44 kilometers northwest of the state capital, Santa Fe, and 82 kilometers north of the state's largest city, Albuquerque. Figure~\ref{fig:study_area} shows the study area and the distribution of reference sites indexed by known species type. The ``Y''-shaped polygon represents a portion of the region known to be dominated by aspen, locust, and oak \citep{coop2016influences, allen2019} where we seek to predict plant cover type. 

Across the Western United States, the last several decades have been marked by increasing forest fire activity with larger proportions burning with high severity than in the past, leading to inhibited recovery mechanisms and even long-term forest conversion of post-fire vegetation \cite{COOP2014}. The fire history of the Jemez Mountains of New Mexico are illustrative of this larger phenomenon with repeated wildfires resulting in a proliferation of large, continuous treeless areas with few remnant live trees. Findings revealed by analyses like ours may provide insight into the ways in which forests in the Western United States are responding to changing climate. 
  
\subsection{Satellite imagery} \label{sec:satellite}
Multispectral imaging, such as the imagery captured by remote sensing satellites, involves measuring the intensity of electromagnetic radiation in narrow bands around specific frequencies/wavelengths \citep{jones2010remote}. The radiation detected by a multispectral satellite and recorded as reflectance intensity is primarily incident radiation from the sun reflecting off the Earth's surface. The amount of energy reflected from these surfaces is usually expressed as a percentage of the total amount of energy striking the objects. A reflectance of $1$ is recorded if all of the light striking a location bounces off and is detected by a radiometric instrument. If none of the light returns from the surface, reflectance is said to be $0$. In most cases, the reflectance intensity value at each location for each interval of the electromagnetic spectrum measured is somewhere between these two extremes.   

When imaging tree or shrub canopies for classification purposes, the amount of radiation that is reflected in the different wavelengths is related to plant chemical properties of the tissue including water, photosynthetic pigments, and structural carbohydrates; the morphology of tree leaves including thickness of cell walls, air spaces, and cuticle wax; and structural traits that influence the signal strength of the member species \citep[i.e., leaf and branch density, leaf angular distribution, canopy coverage, understory vegetation][]{FASS2016}. Reflectance varies across wavelengths and seasons, especially for deciduous species, according to patterns that differ across species, sometimes drastically and sometimes subtly \citep{ASSA2015, ASSA2021}. For example, coniferous and deciduous trees are quite distinct in their respective reflectance patterns, for much the same reason that they are visually distinct to the human eye. In contrast, two deciduous species like quaking aspen and New Mexico locust can be much more difficult to discern at a distance. We hope to leverage publicly available reflectance measurements made across a broad and finely resolved grid of wavelengths to map patterns of vegetation at a level of taxonomic detail approaching individual species.

Reflectance measurements from Sentinel-2 are made at a finite spatial resolution (between 10-60m), representing an average value across areas that may differ in ground cover characteristics and canopy coverage at different locations \cite{FASS2016}. In addition, there are limits to the precision of the instruments used to measure reflectance. Thus, two sites with the same predominant vegetation cover type can be expected to exhibit additional variation in reflectance not attributable to species-specific characteristics. Because it is reasonable to expect species-driven patterns of reflectance to vary slowly in time relative to the revisit frequency, we are able to specify statistical models that isolate this additional variation and uncover consistent patterns that can be used for prediction.

\section{Statistical Model}\label{sec:model}

\subsection{Satellite reflectance}\label{sec:reflectance}
To introduce our proposed hierarchical model for observed reflectance, we first consider a simplified situation in which all locations have a shared vegetation cover type. In this setting, we model each observed reflectance, $r(w, d)$, at wavelength, $w$, on date, $d$, as independent and Gaussian-distributed with mean, $f(w, d)$ and variance $\sigma^2$. We assume that $f(w, d)$ is a continuous function of wavelength and date representing the expected reflectance. Additional variability in observations due to factors such as the finite precision of satellite instrumentation is assumed to be independent of both wavelength and date with magnitude controlled by $\sigma^2$, such that $r(w, d) \sim \mathrm{N}\left(f(w, d), \sigma^2\right)$, where $\mathrm{N}(m, s)$ denotes a normal distribution with mean $m$ and variance $s$.

Different vegetation types are expected to produce different amounts of reflectance for a given wavelength and date owing to biological and ecological differences among plant types (e.g., physical structure, timing of senescence for deciduous species). Indeed, it is these distinguishing characteristics that will provide information needed to predict plant cover type at new locations using satellite imagery. To account for these characteristics, we model reflectances separately for each cover type, and denote by $f_k(w, d)$ the expected reflectance for cover type $k$.

To flexibly and efficiently model the reflectance surface for each cover type, we specify a space of linear functions spanned by the basis, $g_l(w, d), l \in \{1, \dots L\}$, along with cover type-specific coefficients, $\gamma_{lk}$, such that $f_k(w, d) = \sum_{l = 1}^L g_l(w, d) \gamma_{lk}$ \citep{hefley2016}. For notational compactness, let $\boldsymbol{g}(w, d)$ and $\boldsymbol\gamma_k$ denote the vectors of length $L$ containing the evaluation of all $L$ basis functions at $w$ and $d$ and associated coefficients such that the reflectance function can be written as the vector inner product $f_k(w, d) = \boldsymbol{g}^\prime(w, d) \boldsymbol\gamma_k$. In addition, let $\mathbf{\Gamma}$ denote the $L \times K$ matrix who columns are $\boldsymbol\gamma_k$ for $k = 1, \dots, K$.

Because reflectances are by definition constrained to the unit interval, we model the logit-transformed reflectance, which is unbounded ($\mathrm{logit}(z) = \log(\frac{z}{1 - z})$). Hereafter, let $r_{ij}$ denote the $i$th logit-transformed reflectance observation at site $j$ corresponding to wavelength $w_{ij}$ on date $d_{ij}$, let $\boldsymbol{g}_{ij}$ be the length-$L$ vector of basis functions evaluated at $(w_{ij}, d_{ij})$, and let $y_j = k \in \{1, \dots, K\}$ denote which of $K$ vegetation cover types is present at site $j$. We specify the conditional distribution of $r_{ij}$ such that 
\begin{align}
\pi(r_{ij}|y_j = k, \boldsymbol\gamma_k, \sigma^2) &= 
  \mathrm{N}(\boldsymbol{g}_{ij}^\prime\boldsymbol\gamma_{k}, \sigma^2), 
    \quad j \in \{1, \dots, J \}, i \in \{1, \dots, N_j\}, \label{eqn:reflectance}
\end{align}
where $\pi(\cdot)$ denotes a probability density.

\subsection{Cover type}\label{sec:cover type}
Many landscape characteristics are responsible for driving the complex ecological processes that ultimately give rise to the realized patterns of vegetation cover type. For example, many topographical features derived from elevation can help explain the distribution of different cover types (e.g., slope, aspect, heat loading). To learn important relationships between the pattern of observed cover types and these landscape characteristics, and to aid in predicting cover type at new locations, we specify a multinomial logistic model \citep[e.g.,][]{agresti_categorical_2002} for $y_j$ as a function of $P$ site-specific landscape characteristics, $\boldsymbol{x}_j$, and associated coefficients, $\boldsymbol\beta_k$, unique to each cover type, $k \in \{1, \dots, K\}$. Let $\mathbf{X}$ denote the $J \times P$ matrix of covariate values corresponding to landscape characteristics (columns) measured at each site (rows), and let $\mathbf{B}$ denote the $P \times K$ matrix of coefficients for each covariate (rows) and cover type (columns).

In addition, we incorporate spatially smooth random effects that account for residual dependence in cover type for proximal sites. The spatial random effect is defined as a random function over the continuously indexed spatial domain from the linear space spanned by the set of $M$ basis functions, $h_m(\boldsymbol{s}_j), m \in \{1, \dots, M\}$, with cover type-specific coefficients $\boldsymbol\phi_{k}$. We use notation analogous to the basis functions for reflectances and write $\boldsymbol{h}_j$ for the $M$-vector made up of the values of all spatial basis functions evaluated at site $\boldsymbol{s}_j$. The basis functions are specified such that they are linearly independent of, but not necessarily orthogonal to, the landscape characteristics. Let $\mathbf{H}$ denote the $J \times M$ matrix of basis functions (columns) evaluated at each site (rows), and let $\mathbf\Phi$ be the $M \times K$ matrix of associated coefficients.

Let $p_{jk}$ denote the probability that site $j$ has cover type $k$ (i.e., $\mathrm{Pr}(y_j = k) = p_{jk}$) with the constraint that $\sum_{k=1}^K p_{jk} = 1$ for all $j$, and let $\boldsymbol{p}_j$ denote the $K$-vector of probabilities for site $j$. The sum-to-one constraint on probabilities for each site means that not all values in $\mathbf{B}$ and $\mathbf\Phi$ are identifiable. Stated another way, it is not possible to estimate the marginal effects of each covariate and spatial basis function on the probability for each cover type, only the effect relative to a baseline cover type. By convention \citep[e.g.,][]{agresti_categorical_2002}, we define $\boldsymbol\beta_K$ and $\boldsymbol\phi_K$, the coefficients associated with level $K$, to be 0. This leads to the following formula for $p_{jk}$, the probability of observing cover type $k$ at site $j$:
\begin{align}
\log(\eta_{jk}) &=
  \boldsymbol{x}^\prime_j \boldsymbol\beta_k + \boldsymbol{h}^\prime_{j} \boldsymbol\phi_k,
  \quad k \in \{1, \dots, K\} \label{eqn:eta} \\
p_{jk} &= \frac{\eta_{jk}}{1 + \sum_{k = 1}^{K - 1} \eta_{jk}} \label{eqn:norm}.
\end{align}
The intermediate quantities $\eta_{jk}$ highlight the construction of each probability in a way that respects the natural constraints on $p_{jk}$, yet admits a simple linear framework. In particular, the logarithm on the left hand side of \eqref{eqn:eta} ensures that $\eta_{jk}$ is always positive, and the normalization in \eqref{eqn:norm} ensures the sum-to-one constrains for probabilities at each site. A graphical representation of the complete hierarchical model as a directed acyclic graph is available in Supplementary Materials Figure~\ref{fig:DAG}.

\section{Implementation}\label{implementation}

\subsection{Bayesian inference and prediction}
We implement our model within a Bayesian paradigm, specifying prior beliefs for all unknown parameters through carefully chosen probability distributions and basing inference and predictions on the resulting joint posterior distribution. An analytical form for the posterior is not available, so we rely instead on sampling from the posterior distribution using Markov chain Monte Carlo (MCMC) techniques, facilitated by the R package NIMBLE \citep{NIMBLE, NIMBLEJCGS}.

We specify a mean-zero multivariate normal distribution with covariance $\mathbf{\Sigma}_\gamma~=~\rho~\mathbf{J}_K~+~(1~-~\rho)\mathbf{I}_K$ for all $l \in \{1, \dots, L\}$, where $\boldsymbol\gamma_l$ represents the $K$ species-specific coefficients associated with the $l$th basis function, $\boldsymbol{g}_l$ (note: $\mathbf{I}_K$ denotes the $K \times K$ identity matrix, and $\mathbf{J}_K$ denotes the $K \times K$ matrix of ones). We set $\rho = 0.9$ corresponding to our belief that reflectances for each of the three vegetation types are likely to be very similar.

We specify mean-zero, independent normal distributions for all coefficients in $\mathbf\Phi$ and $\mathbf{B}$. The relative variance of the coefficients for the spatial random effect reflects our belief that the effects of the landscape characteristics on the distribution of each vegetation type will be much larger than the residual variation. Finally, we specify a Gamma distribution prior for $\sigma^2$. The shape and scale hyper-parameters were chosen based on an exploration of simulated data from the prior predictive distribution \citep{Gabry2019}, and we note that the posterior of $\sigma^2$ is quite insensitive to this choice given the volume of reflectance data in our application.

\subsubsection{Marginalized model for reflectance}
For sites where cover type is not measured (i.e., only reflectance and landscape characteristics are available), $y_j$ can be integrated out of the hierarchical model to produce a mixture distribution. Let $\mathcal{J}_\mathrm{obs} = \{j: y_j \text{ observed}\}$ be the set of indices corresponding to sites where cover type has been measured with size $|\mathcal{J}_\mathrm{obs}| = J_\mathrm{obs}$, and let $\mathcal{J}^\mathrm{c}_\mathrm{obs} = \{1, \dots, J\} \setminus \mathcal{J}_\mathrm{obs}$ be the complementary set of sites where cover type has not been measured. The conditional distribution of reflectance given $\boldsymbol{p}_{j}$, rather than $y_j$, is 
\begin{align}
\pi\left(r_{ij}|\boldsymbol{p}_j, \boldsymbol\gamma_k, \sigma^2\right) &= 
  \sum_{k = 1}^K p_{jk} \mathrm{N}\!\left(
    \boldsymbol{g}_{ij}^\prime\boldsymbol\gamma_{k}, \sigma^2 
  \right), \quad j \in \mathcal{J}_\mathrm{obs}^\mathrm{c}, i \in \{1, \dots, N_j\}.
\end{align}
The primary benefit of the marginalized model is that model fitting can proceed without having to update the latent cover type at every site where it has not been measured, which improves the efficiency of our MCMC algorithm. The model is otherwise equivalent to the hierarchical version, and samples from the posterior distributions for $y_j$ can be obtained via composition sampling. Where cover type is available, it may still be used, yielding a final likelihood that is partitioned into two components as 
\begin{align}
L(\boldsymbol\theta | \boldsymbol{r}, \boldsymbol{y}_\mathrm{obs}) &= 
  \left(\prod_{j\in\mathcal{J}_\mathrm{obs}} \prod_{i = 1}^{N_j} 
    \mathrm{N}\!\left(\boldsymbol{g}_{ij}^\prime\boldsymbol\gamma_{y_j}, \sigma^2 \right)\right)
  \left(\prod_{j\in\mathcal{J}^\mathrm{c}_\mathrm{obs}} \prod_{i = 1}^{N_j}
    \sum_{k = 1}^K p_{jk} \mathrm{N}\!\left(\boldsymbol{g}_{ij}^\prime\boldsymbol\gamma_{k}, \sigma^2\right)\right),
\end{align}
where $\gamma_{y_j}$ is used as shorthand notation for $\sum_{k = 1}^K \boldsymbol\gamma_k \boldsymbol{1}_{y_j = k}$ and $\boldsymbol\theta$ represents the collection of unknown model parameters $\{\mathbf\Gamma, \mathbf\Phi, \mathbf{B}, \sigma^2\}$.

\subsubsection{Interpreting the effects of landscape characteristics on cover type}
One of the primary scientific interests motivating our approach is inference about the relative effects of various landscape characteristics on vegetation cover type. The relevant model parameters needed to answer such questions are primarily $\mathbf{B}$ and, secondarily, $\mathbf\Phi$. The former captures the direct effects of measured landscape predictors, $\mathbf{X}$, and the latter residual spatial patterns in cover type not fully explained by $\mathbf{X}$. 

While it is straightforward to interpret the posterior of a particular $\beta_{pk}$ up to its sign (i.e., positive values represent an increase in the probability of finding species $k$ for an increase in $x_p$ relative to the baseline species $K$), the nonlinearity of the effects induced by the constraints that $0 \leq p_{jk} \leq 1$ and $\sum_{k=1}^K p_{jk} = 1$ limit the intuitive value of the raw coefficients. Instead, motivated by \cite{SCHARF2022}, we propose a visual summary of the effects using marginal probability curves as functions of each covariate (e.g., Figure~\ref{fig:veg_prob_curves}).

\subsubsection{Suitability vs. presence} \label{sec:suitability}
The posterior distribution of $y_j$ represents beliefs about which cover type is most likely to exist at site $j$ given all available information including landscape characteristics and observed reflectances. The related posterior distribution of $\boldsymbol{p}_j$ represents a distinct, but nevertheless useful quantity. It differs from the posterior of $y_j$ in the way it involves local reflectance measurements. Namely, the observations $\boldsymbol{r}_j$ only inform the posterior of $\boldsymbol{p}_j$ through their influence on $\mathbf{B}$ and $\mathbf\Phi$, and not through their compatibility with the inferred reflectance surface. Thus, the posterior of $\boldsymbol{p}_j$ represents what we believe the probabilities for each cover type should be at sites like $j$ if we only observe the local landscape characteristics, and might therefore be best understood as a site's ``suitability'' for each vegetation type \citep[see][for an analogous example involving patterns of climate sensitivity/robustness in plant communities]{SCHA2021}. We compare summaries of each in our application to the distribution of post-fire resprouting deciduous species in the Jemez mountains of New Mexico (Section~\ref{sec:application}). Samples from the posterior can be obtained through composition sampling, by first drawing $\boldsymbol\theta^{(t)}, t = 1, \dots, T$ from the posterior, and then computing $p_{jk}(\boldsymbol\theta^{(t)})$ for each site and cover type using \eqref{eqn:eta} and \eqref{eqn:norm}.

\subsection{Computational challenge}\label{sec:computational-challenge}
We developed our proposed model for reflectance mindful of computational demands associated with its implementation. The number of reflectance measurements in our motivating application is large, totaling over 50,000 at the 60 sites where we have observations of cover type, and over 17,000,000 across all sites in the study area (see Supplementary Materials Table~\ref{tab:constants}), which is itself a subset of the larger region of interest. We specify the functional forms of the spatial random effect in the model for cover type probabilities, $p_{jk}$, and the reflectance surface, $f(w_{ij}, d_{ij})$, using linear combinations of basis functions, which leads to conditional independence in the relevant components of the hierarchical models and substantial computational savings relative to, for example, covariance function-based Gaussian processes \citep{hefley2016}. Even though these model features mean that computational demands only grow linearly in the number of reflectance observations, the computational cost is nevertheless substantial. In our implementation using slice sampling via NIMBLE, obtaining sufficient samples from the posterior distribution for reliable inference required approximately 15 hours when limiting attention to the 60 primary sites and 20 additional sites. The extra time required to include an additional 40 sites was approximately 10 hours. In addition, demands on memory grew to exceed 20GB for an additional 420 sites. Thus, it was not feasible to implement our proposed model to estimate the joint posterior distribution of all model parameters based all available reflectance data using broadly available computing resources.

There are two primary goals driving our analysis. First, we are interested in understanding the relationship between the measured landscape characteristics and the distribution of the three focal species of deciduous plants. To this end, we are interested in obtaining posterior distributions for $\mathbf{B}$ and $\boldsymbol\Phi$ that make use of as much available information as possible to minimize posterior uncertainty. As we show in the simulation study and application, beliefs about these parameters continue to be refined as more and more sites' worth of information is included, and thus, we pursue posteriors based on as many reflectance measurements as our computational resources and time allow.

Second, we are interested in predicting cover type at new locations, where we assume additional reflectance measurements have been made. We demonstrate in the following section how to efficiently utilize all available data to achieve accurate cover type predictions. 

\subsubsection{Predictions at new sites}
The following conditional independence relationship follows from the hierarchical model structure:
\begin{align}
\boldsymbol{y}_\mathcal{J} | \boldsymbol\theta &\perp\!\!\!\perp 
  \boldsymbol{r}_{\mathcal{J}^\mathrm{c}}, \boldsymbol{y}_{\mathcal{J}^\mathrm{c}} 
  \; \forall \; \mathcal{J} \subseteq \{1, \dots, J \}. \label{eqn:ci}
\end{align}
Let $\boldsymbol{y}_\mathrm{obs}^\mathrm{c} = \{y_j: j \in \mathcal{J}_\mathrm{obs}^\mathrm{c}\}$ represent cover type at locations where it has not already been observed. The target posterior for predicting cover type given observed cover types, $\boldsymbol{y}_\mathrm{obs}$, and associated reflectances, $\boldsymbol{r}_\mathrm{obs}$, as well as reflectances recorded where cover type has not been observed, $\boldsymbol{r}_\mathrm{obs}^\mathrm{c} = \{r_{ij}: j \in \mathcal{J}_\mathrm{obs}^\mathrm{c}\}$, is
\begin{align}
\pi(\boldsymbol{y}_\mathrm{obs}^\mathrm{c} | \boldsymbol{y}_\mathrm{obs}, 
  \boldsymbol{r}_\mathrm{obs}, \boldsymbol{r}_\mathrm{obs}^\mathrm{c})
  &= \mathrm{E}_{\boldsymbol\theta | \boldsymbol{y}_\mathrm{obs}, 
      \boldsymbol{r}_\mathrm{obs}, \boldsymbol{r}_\mathrm{obs}^\mathrm{c}}
      \big[\pi(\boldsymbol{y}_\mathrm{obs}^\mathrm{c} | \boldsymbol{r}_\mathrm{obs}^\mathrm{c}, \boldsymbol\theta)\big], 
      \label{eqn:expect}
\end{align}
where we have used \eqref{eqn:ci} with $\mathcal{J} = \mathcal{J}_\mathrm{obs}^\mathrm{c}$ and we note that $\pi(\boldsymbol{y}_\mathrm{obs}^\mathrm{c} | \boldsymbol{r}_\mathrm{obs}^\mathrm{c}, \boldsymbol\theta)$ is treated as a function of $\boldsymbol\theta$ in the expectation. The probability density $\pi(\boldsymbol{y}_\mathrm{obs}^\mathrm{c} | \boldsymbol{r}_\mathrm{obs}^\mathrm{c}, \boldsymbol\theta)$ is discrete without a compact parametric form. In principle, it can be evaluated numerically via
\begin{align}
\pi(\boldsymbol{y}_\mathrm{obs}^\mathrm{c} | \boldsymbol{r}_\mathrm{obs}^\mathrm{c}, \boldsymbol\theta ) &= 
  \frac{\pi(\boldsymbol{r}_\mathrm{obs}^\mathrm{c} | \boldsymbol\theta, \boldsymbol{y}_\mathrm{obs}^\mathrm{c})
        \pi(\boldsymbol{y}_\mathrm{obs}^\mathrm{c} | \boldsymbol\theta)}
      {\sum_{\boldsymbol{y} \in \{1, \dots, K\}^{J_\mathrm{obs}^\mathrm{c}}} 
        \pi(\boldsymbol{r}_\mathrm{obs}^\mathrm{c} | \boldsymbol\theta, \boldsymbol{y}_\mathrm{obs}^\mathrm{c} = \boldsymbol{y})
        \pi(\boldsymbol{y}_\mathrm{obs}^\mathrm{c} = \boldsymbol{y} | \boldsymbol\theta)} \label{eqn:ygivenrtheta}
\end{align}
where $\{1, \dots, K\}^{J_\mathrm{obs}^\mathrm{c}}$ is the entire sample space for $\boldsymbol{y}_\mathrm{obs}^\mathrm{c}$ (see Sections~\ref{sec:reflectance} and \ref{sec:cover type} for definitions of $\pi(\boldsymbol{r}_\mathrm{obs}^\mathrm{c} | \mathbf\Gamma, \sigma^2, \boldsymbol{y}_\mathrm{obs}^\mathrm{c})$ and $\pi(\boldsymbol{y}_\mathrm{obs}^\mathrm{c} | \mathbf{B}, \mathbf\Phi)$, respectively).

Equations \eqref{eqn:expect} and \eqref{eqn:ygivenrtheta} suggest a Monte Carlo approach for estimating the target distribution using samples from the posterior $\pi(\boldsymbol\theta | \boldsymbol{y}_\mathrm{obs}, \boldsymbol{r}_\mathrm{obs}, \boldsymbol{r}_\mathrm{obs}^\mathrm{c})$.  (see Algorithm~\ref{alg:pred}). However, this is impractical for two reasons. First, samples from $\pi(\boldsymbol\theta | \boldsymbol{y}_\mathrm{obs}, \boldsymbol{r}_\mathrm{obs}, \boldsymbol{r}_\mathrm{obs}^\mathrm{c})$ are difficult to obtain because the sheer size of $\boldsymbol{r}_\mathrm{obs}^\mathrm{c}$ makes likelihood calculations prohibitively time consuming. Second, as $J_\mathrm{obs}^\mathrm{c}$ grows, the size of the sample space $\{1, \dots, K\}^{J_\mathrm{obs}^\mathrm{c}}$ increases exponentially, making it infeasible to evaluate the denominator in \eqref{eqn:ygivenrtheta} for all but very modest numbers of new sites. We address each of these limitations with an approximate method for posterior inference.

\begin{algorithm}[ht]
\caption{Monte Carlo estimate for posterior of cover type}
\label{alg:pred}
\begin{algorithmic}[1]
  \State Define $u(\boldsymbol\theta, \boldsymbol{y}_\mathrm{obs}^\mathrm{c}) = 
    \frac{\pi(\boldsymbol{r}_\mathrm{obs}^\mathrm{c} | \boldsymbol\theta, \boldsymbol{y}_\mathrm{obs}^\mathrm{c})
        \pi(\boldsymbol{y}_\mathrm{obs}^\mathrm{c} | \boldsymbol\theta)}
      {\sum_{\boldsymbol{y} \in \{1, \dots, K\}^{J_\mathrm{obs}^\mathrm{c}}} 
        \pi(\boldsymbol{r}_\mathrm{obs}^\mathrm{c} | \boldsymbol\theta, \boldsymbol{y}_\mathrm{obs}^\mathrm{c} = \boldsymbol{y})
        \pi(\boldsymbol{y}_\mathrm{obs}^\mathrm{c} = \boldsymbol{y} | \boldsymbol\theta)}$
  \For{t = 1, \dots, T}
    \State Draw $\boldsymbol\theta^{(t)} \sim 
      \pi(\boldsymbol\theta | \boldsymbol{y}_\mathrm{obs}, 
      \boldsymbol{r}_\mathrm{obs}, \boldsymbol{r}_\mathrm{obs}^\mathrm{c})$
    \State Evaluate $\pi(\boldsymbol{r}_\mathrm{obs}^\mathrm{c} | \boldsymbol\theta^{(t)}, 
      \boldsymbol{y}_\mathrm{obs}^\mathrm{c})\pi(\boldsymbol{y}_\mathrm{obs}^\mathrm{c} | \boldsymbol\theta^{(t)}) \;\forall\;
      \boldsymbol{y}_\mathrm{obs}^\mathrm{c} \in \{1, \dots, K\}^{J_\mathrm{obs}^\mathrm{c}}$
    \State Compute $u(\boldsymbol\theta^{(t)}, \boldsymbol{y}_\mathrm{obs}^2)$
  \EndFor
  \Return $\frac{1}{M}\sum_{m = 1}^M u(\boldsymbol\theta^{(t)}, \boldsymbol{y}_\mathrm{obs}^\mathrm{c})
  \approx \pi(\boldsymbol{y}_\mathrm{obs}^\mathrm{c} | \boldsymbol{y}_\mathrm{obs},
    \boldsymbol{r}_\mathrm{obs}, \boldsymbol{r}_\mathrm{obs}^\mathrm{c})$
\end{algorithmic}
\end{algorithm}

To address the issue with sampling from the full target posterior, we propose using samples from $\pi(\boldsymbol\theta | \boldsymbol{y}_\mathrm{obs}, \boldsymbol{r}_\mathrm{obs}, \boldsymbol{r}_\mathrm{add})$, where $\boldsymbol{r}_\mathrm{add} \subset \boldsymbol{r}_\mathrm{obs}^\mathrm{c}$ are logit-transformed reflectance values measured at a subset of the sites where new predictions are desired. Let $\mathcal{J}_\mathrm{add} \subset \mathcal{J}_\mathrm{obs}^\mathrm{c}$ with $|\mathcal{J}_\mathrm{add}| = J_\mathrm{add} << J_\mathrm{obs}^\mathrm{c}$, and $\boldsymbol{r}_\mathrm{add} = \{r_{ij}: j \in \mathcal{J}_\mathrm{add}\}$. Essentially, we propose fitting the model to as many additional reflectance measurements as is computationally feasible, along with those associated with with observed cover types. We note that the above definition for $\boldsymbol{r}_\mathrm{add}$ is based on site indices, and thus reflectances across spectra and season for a particular site are either all included, or all omitted from the analysis. More sophisticated sampling designs may offer improvements in prediction, but are beyond the scope of the present work.

To address the issue concerning the size of the sample space for $\boldsymbol{y}_\mathrm{obs}^\mathrm{c}$, we propose limiting attention to the marginal distributions $\pi(y_j | \boldsymbol{y}_\mathrm{obs}, \boldsymbol{r}_\mathrm{obs}, \boldsymbol{r}_\mathrm{obs}^\mathrm{c})$, $j \in \mathcal{J}_\mathrm{obs}^\mathrm{c}$, thereby forgoing joint information about cover type in exchange for computational efficiency. Combining the two proposed adjustments, the new target distributions are 
$\mathrm{E}_{\boldsymbol\theta | \boldsymbol{y}_\mathrm{obs}, 
  \boldsymbol{r}_\mathrm{obs}, \boldsymbol{r}_\mathrm{add}} 
\big[\pi(y_j | \boldsymbol{r}_j, \boldsymbol\theta) \big], 
  \; j \in \mathcal{J}_\mathrm{obs}^\mathrm{c}$.
Two cases emerge, depending on whether a new site is inside or outside the set $\mathcal{J}_\mathrm{add}$. For $j \in \mathcal{J}_\mathrm{add}$, 
\begin{align}
  \mathrm{E}_{\boldsymbol\theta | \boldsymbol{y}_\mathrm{obs}, \boldsymbol{r}_\mathrm{obs}, \boldsymbol{r}_\mathrm{add}} 
    \big[\pi(y_j | \boldsymbol{r}_j, \boldsymbol\theta) \big]
  &= \mathrm{E}_{\boldsymbol\theta | \boldsymbol{y}_\mathrm{obs}, \boldsymbol{r}_\mathrm{obs}, \boldsymbol{r}_\mathrm{add}} 
    \big[\pi(y_j | \boldsymbol{r}_\mathrm{add}, \boldsymbol\theta) \big] \label{eqn:case1}\\
  &= \pi(y_j | \boldsymbol{y}_\mathrm{obs}, \boldsymbol{r}_\mathrm{obs}, \boldsymbol{r}_\mathrm{add}).
\end{align}
Thus, for sites associated with additional, included reflectance measurements, we obtain the exact marginal posterior of cover type based on observed and additional sites.

When $j \in \mathcal{J}_\mathrm{obs}^\mathrm{c} \setminus \mathcal{J}_\mathrm{add}$, then 
$\mathrm{E}_{\boldsymbol\theta | \boldsymbol{y}_\mathrm{obs}, \boldsymbol{r}_\mathrm{obs}, \boldsymbol{r}_\mathrm{add}} 
  \big[\pi(y_j | \boldsymbol{r}_j, \boldsymbol\theta) \big]$ 
cannot be written in the form of a posterior because $\boldsymbol{r}_j \not\subseteq \boldsymbol{r}_\mathrm{add}$, and thus there is a ``mismatch'' between the argument and random variable associated with the expectation (i.e., $\boldsymbol{r}_j$ does not appear in the subscript). However, if $\boldsymbol\theta | \boldsymbol{y}_\mathrm{obs}, \boldsymbol{r}_\mathrm{obs}, \boldsymbol{r}_\mathrm{add}$ and $\boldsymbol\theta | \boldsymbol{y}_\mathrm{obs}, \boldsymbol{r}_\mathrm{obs}, \boldsymbol{r}_\mathrm{add}, \boldsymbol{r}_\mathrm{j}$ have approximately the same distribution, then the continuous mapping theorem \citep[][Ch. 5]{billingsley_probability_1995} implies that 
\begin{align}
\mathrm{E}_{\boldsymbol\theta | \boldsymbol{y}_\mathrm{obs}, \boldsymbol{r}_\mathrm{obs}, \boldsymbol{r}_\mathrm{add}} 
  \big[\pi(y_j | \boldsymbol{r}_j, \boldsymbol\theta) \big] 
&\approx \mathrm{E}_{\boldsymbol\theta | \boldsymbol{y}_\mathrm{obs}, \boldsymbol{r}_\mathrm{obs}, \boldsymbol{r}_\mathrm{add}, \boldsymbol{r}_j}
  \big[\pi(y_j | \boldsymbol{r}_j, \boldsymbol\theta) \big] \\
&= \pi(y_j | \boldsymbol{y}_\mathrm{obs}, \boldsymbol{r}_\mathrm{obs}, \boldsymbol{r}_\mathrm{add}, \boldsymbol{r}_j).
\end{align}
Thus, for sites outside of $\mathcal{J}_\mathrm{add}$, we obtain a distribution approximately equal to the marginal posterior of cover type based on data from all sites in $\mathcal{J}_\mathrm{obs}\cup\mathcal{J}_\mathrm{add}\cup j$.

We note that while posterior inference about the model parameters $\boldsymbol\theta$ is based on a strict subset of the measured reflectance values (i.e., $\boldsymbol{r}_\mathrm{obs}, \boldsymbol{r}_\mathrm{add}$), \textit{all} reflectance values ultimately contribute to predictions about cover type. Algorithm~\ref{alg:pred_approx} provides a practical alternative to Algorithm~\ref{alg:pred} for estimating marginal posteriors for cover type that we use in all subsequent analyses.

\begin{algorithm}[ht]
\caption{Monte Carlo estimate for \textit{approximate marginal} posteriors of cover type}
\label{alg:pred_approx}
\begin{algorithmic}[1]
  \State Define $u(\boldsymbol\theta, y_j) 
    = \frac{\pi(\boldsymbol{r}_j | \boldsymbol\theta, y_j)
        \pi(y_j | \boldsymbol\theta)}{\sum_{y \in \{1, \dots, K\}} 
        \pi(\boldsymbol{r}_j | \boldsymbol\theta, y_j = y)
        \pi(y_j = y | \boldsymbol\theta)}$
  \For{t = 1, \dots, T}
    \State Draw $\boldsymbol\theta^{(t)} \sim 
      \pi(\boldsymbol\theta | \boldsymbol{y}_\mathrm{obs}, 
      \boldsymbol{r}_\mathrm{obs}, \boldsymbol{r}_\mathrm{add})$
    \For{$j \in \mathcal{J}_\mathrm{obs}^\mathrm{c}$}
      \State Evaluate $\pi(\boldsymbol{r}_j | \boldsymbol\theta^{(t)}, y_j = y)
        \pi(y_j = y | \boldsymbol\theta^{(t)}) \;\forall\; y \in \{1, \dots, K\}$
      \State Compute $u(\boldsymbol\theta^{(t)}, y_j)$
    \EndFor
  \EndFor
  \State \Return $\frac{1}{M}\sum_{m = 1}^M u(\boldsymbol\theta^{(t)}, y_j)
  \approx \pi(y_j | \boldsymbol{y}_\mathrm{obs}, 
    \boldsymbol{r}_\mathrm{add}, \boldsymbol{r}_j), \;
    j \in \mathcal{J}_\mathrm{obs}^\mathrm{c}$
\end{algorithmic}
\end{algorithm}

\subsubsection{Smoothly changing posteriors}\label{sec:smoothly_changing}

There are two reasons why we expect that the posterior distributions for $\boldsymbol\theta$ based on data sets that differ only by $\boldsymbol{r}_j$ could be approximately equal. First, the amount of information contained in $\boldsymbol{r}_j$ about $\mathbf{B}$ and $\boldsymbol\Phi$ will be small, because the values are associated with a single site. Thus, even if there is rich information about which cover type gave rise to $\boldsymbol{r}_j$, it will contribute only one additional observation to learn about the spatial characteristics of the underlying process.

Second, in situations such as our motivating application to vegetation cover type, $\boldsymbol{r}_\mathrm{obs}$ and $\boldsymbol{r}_\mathrm{add}$ will typically include a very large number of observations ($\boldsymbol{r}_\mathrm{obs}$ alone includes 53,688 in our application), and thus the amount of \textit{new} information contained about the reflectance surfaces (i.e., $\mathbf\Gamma$) in $\boldsymbol{r}_j$ will be small. In other words, we expect that the marginal ``pen-ultimate'' posterior, $\boldsymbol\Gamma | \boldsymbol{y}_\mathrm{obs}, \boldsymbol{r}_\mathrm{obs}, \boldsymbol{r}_\mathrm{add}$, will already be highly concentrated, even without considering $\boldsymbol{r}_\mathrm{j}$. We evaluate these two claims in the context of both a simulation study (Section~\ref{sec:simulation-study}) and our application to the spatial distribution of deciduous vegetation in the Jemez mountains of New Mexico (Section~\ref{sec:application}).

\section{Simulation study}\label{sec:simulation-study}
\subsection{Simulating a hypothetical landscape and reflectances}
We conducted a simulation study to investigate the practical value of our proposed model and implementation plan. Our simulation study was intended to closely mirror important characteristics of our motivating application, scaled-down to a manageable size that allowed us to explore the impacts of implementing the model fitting approach for different $J_\mathrm{add}$.

\begin{figure}[H]
\includegraphics[width=\linewidth]{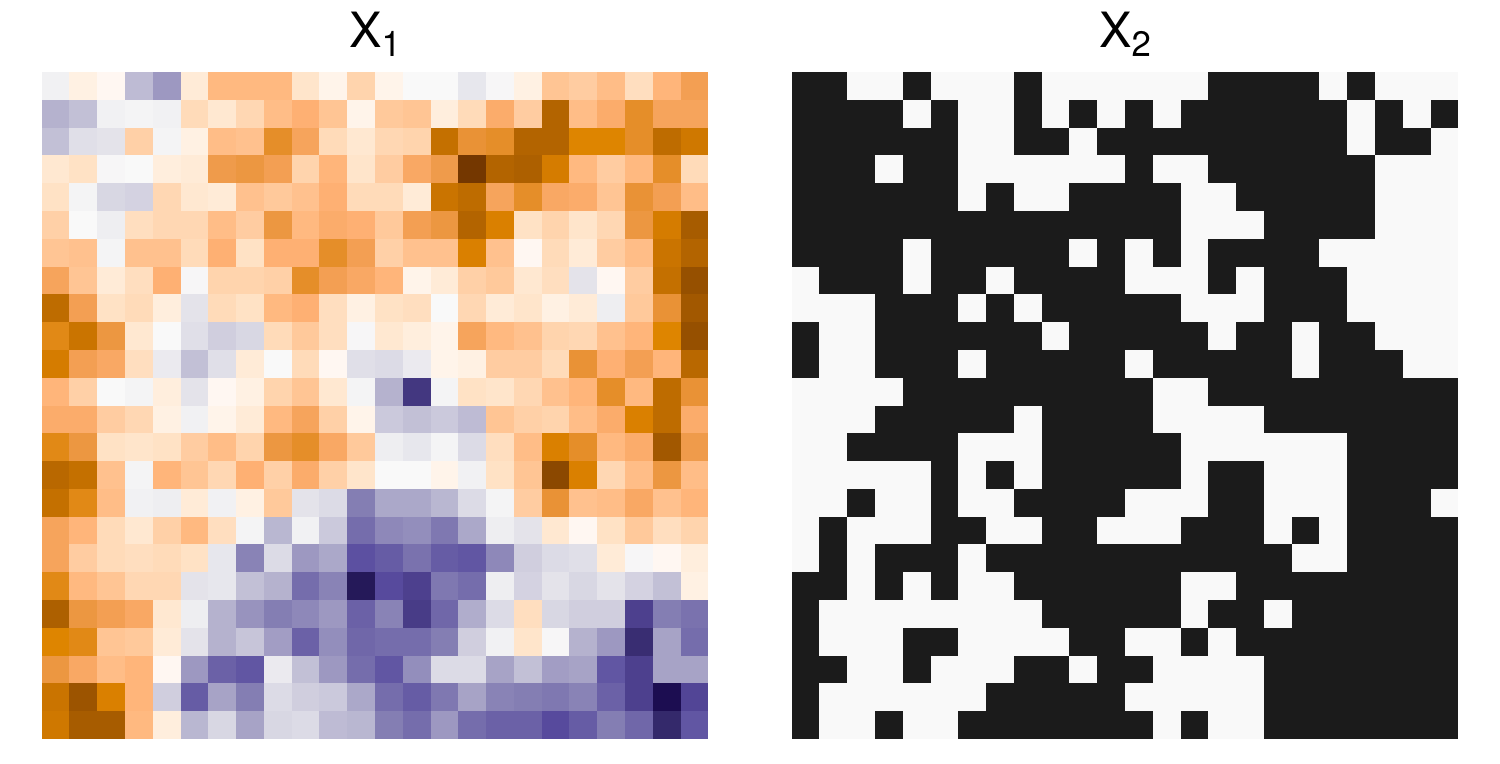}
\caption{Synthetic covariates used in simulation study. $X_1$ is a continuous predictor drawn from a Gaussian process with exponential covariance function, and $X_2$ is binary variable created from thresholding another realization. Covariates were constructed to contain positive spatial autocorrelation to mimck realistic landscape characterististics.}
\label{fig:sim_cov}
\end{figure}

To simulate cover type, we created basis functions and synthetic landscape covariates over a square $24 \times 24$ regular grid on the unit square, $[0, 1] \times [0, 1]$, simulated values of $\mathbf{B}$ and $\mathbf{\Phi}$ from the same prior distributions used in the application to vegetation cover (Supplementary Materials Table~\ref{tab:priors}), and then simulated cover types according to \eqref{eqn:eta}--\eqref{eqn:norm}. Synthetic landscape covariates included an intercept, a centered and scaled realization from a Gaussian process with exponential covariance function, and a binary variable constructed by thresholding another realization from a Gaussian process with exponential covariance function (Supplementary Materials Figure~\ref{fig:sim_cov}). Spatial dependence was included in covariates in an effort to produce realistic predictors mimicking commonly used landscape characteristics such as elevation and slope.

Supplementary Materials Figure~\ref{fig:sim_prob_known} shows the linear effects $\boldsymbol{x}_j^\prime \boldsymbol\beta_k$ (top row) and resulting marginal probabilities (bottom row) of a site having a given cover type as functions of the two synthetic covariates. Each curve represents the probability of observing a particular cover type for the 36 sites in $\mathcal{J}_\mathrm{obs}$. The variation among curves depicts the spatial random effect given by $\boldsymbol{h}_j^\prime\boldsymbol\phi_k$. Note that the linear and spatial effects for cover type $k = 3$ are fixed to be zero to ensure identifiability (top row), yet the resulting probability curves vary for all cover types (bottom row). Supplementary Materials Figure~\ref{fig:sim_spatial_probs} depicts the combined log-linear covariate and spatial effects at each site (top row) as well as the resulting marginal cover type probabilities (bottom row). Of the 576 total locations, $J_\mathrm{obs} = 36$ were chosen completely at random to represent locations where cover type was observed, under the constraint that $\boldsymbol{y}_\mathrm{obs}$ included 12 cases of each cover type. Supplementary Materials Figure~\ref{fig:sim_cover} shows the realized pattern of cover types with darker pixels corresponding to those randomly selected to represent sites with recorded cover type. 

To simulate reflectance values associated with each site, we created additional basis functions over a regular grid of 6 wavelengths and 12 dates, simulated values for $\mathbf{\Gamma}$ and $\sigma^2$ from their respective prior distributions, and then simulated reflectances according to \eqref{eqn:reflectance}. Basis functions for both the reflectance surfaces and spatial random effects were constructed using thin-plate splines of degree $L = M = 30$ over a regular grid of knots. The solid lines in Supplementary Materials Figure~\ref{fig:sim_reflectance} show profiles of the simulated reflectance surface in time for each of the 6 observed wavelengths. Similar patterns in the reflectance profiles arise from our prior specification of strong positive dependence in the coefficients, $\mathbf\Gamma$, corresponding to three species with similar phenotypical characteristics.

\subsection{Model fitting and inference}
We implemented our Bayesian model fitting procedure for randomly selected locations $\mathcal{J}_\mathrm{add}$ of sizes $J_\mathrm{add}$ starting at 20 and incrementing by 40 up to $J_\mathrm{obs}^\mathrm{c} = 576 - 36 = 540$. Slice sampling MCMC algorithms were created using the NIMBLE package \citep{NIMBLE, NIMBLEJCGS} for the R statistical computing environment \citep{R}, and 1,000 posterior samples retained after an initial burnin of 1,000 iterations. Visual inspection of trace plots showed no evidence of failure to converge, and effective sample size calculations showed sufficient values for adequate representation of posterior distributions for all parameters of interest. See Supplementary Materials for additional details.

Figure~\ref{fig:beta_sim} shows posterior summaries of $\mathbf{B}$ in which each plot corresponds to one of $P = 3$ covariate effects. Within each plot there are two collections of boxplots (left, right) corresponding to $k = 1, 2$, respectively. Each collection of boxplots shows the posterior $\pi(\beta_{pk} | \boldsymbol{y}_\mathrm{obs}, \boldsymbol{r}_\mathrm{obs}, \boldsymbol{r}_\mathrm{add})$ for values of $J_\mathrm{add}$ that increase from 20 to 540 moving from left to right. The horizontal red line shows the value of $\beta_{pk}$ used to simulate the data. As expected, posteriors for all effects generally contract as $J_\mathrm{add}$ increases. The posterior median and quantiles change smoothly as a function of $J_\mathrm{add}$, supporting the first hypothesis made in Section~\ref{sec:smoothly_changing} that posteriors differing by a small number of sites will be approximately equal. 

\begin{figure}[ht]
\includegraphics[width=\linewidth]{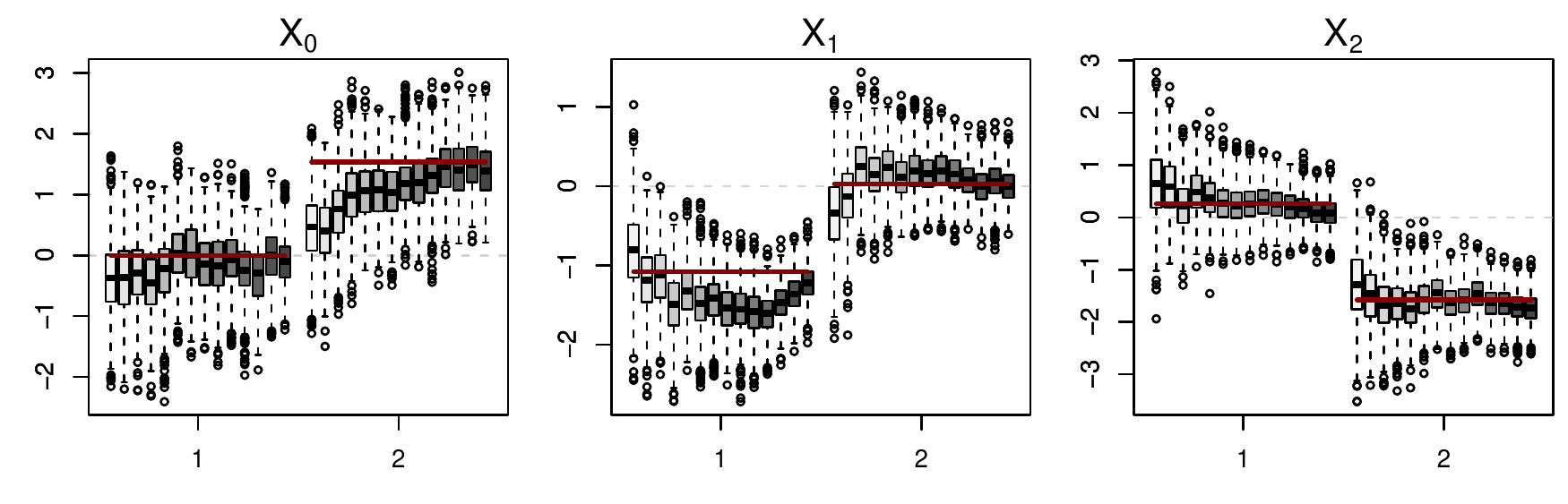}
\caption{Boxplots of marginal posterior distributions for $\mathbf{B}$ across increasing sample sizes. Each subplot corresponds to one of $P = 3$ landscape predictors. Within each subplot, boxplots are grouped by cover type species, with the values for the baseline cover type, k = 3, taken to be 0.}
\label{fig:beta_sim}
\end{figure}

Though the posterior of $\mathbf{B}$ changes gradually for small changes in the size of $J_\mathrm{add}$, it can change substantially when hundreds of additional sites' reflectance measurements are included. In particular, we found that posterior distributions for marginal probability functions for each predictor were substantially contracted for $J_\mathrm{add} = 540$ compared to $20$ (see Supplementary Materials Figure~\ref{fig:sim_prob}).

\begin{figure}[H]
\begin{subfigure}[b]{\linewidth}
  \includegraphics[width=\linewidth]{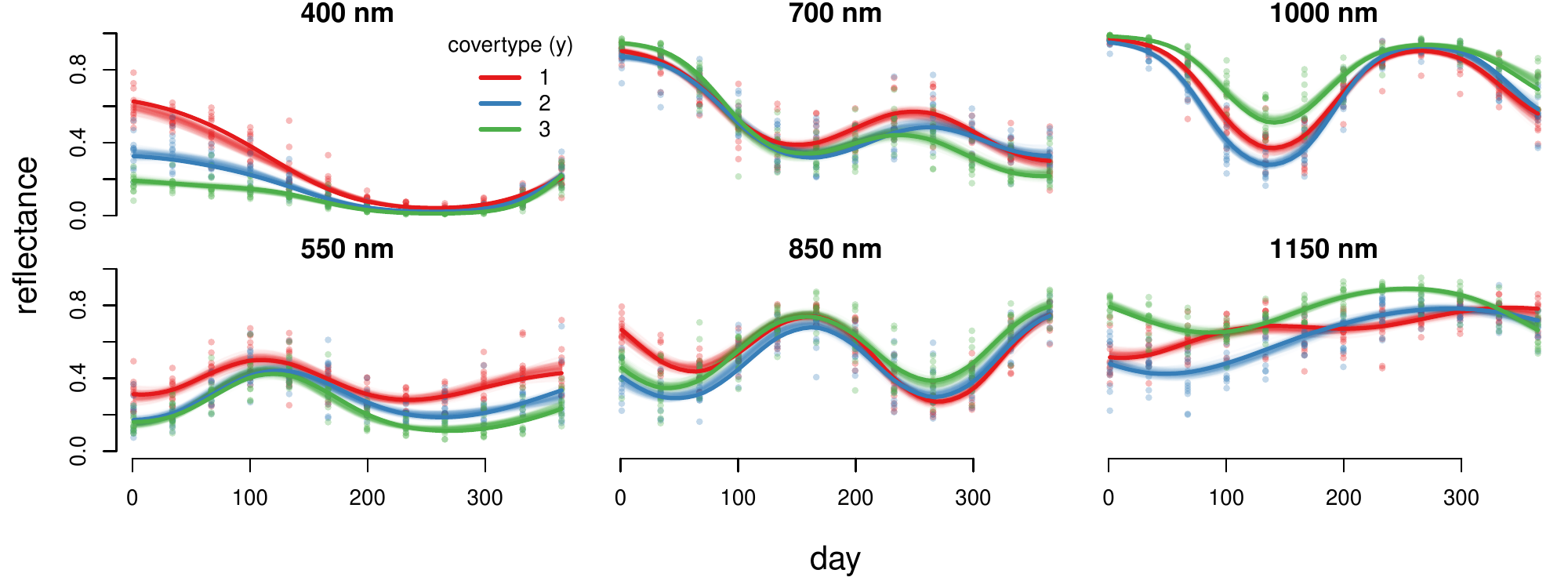}
  \label{fig:reflectance_20}
\end{subfigure}

\begin{subfigure}[b]{\linewidth}
  \includegraphics[width=\linewidth]{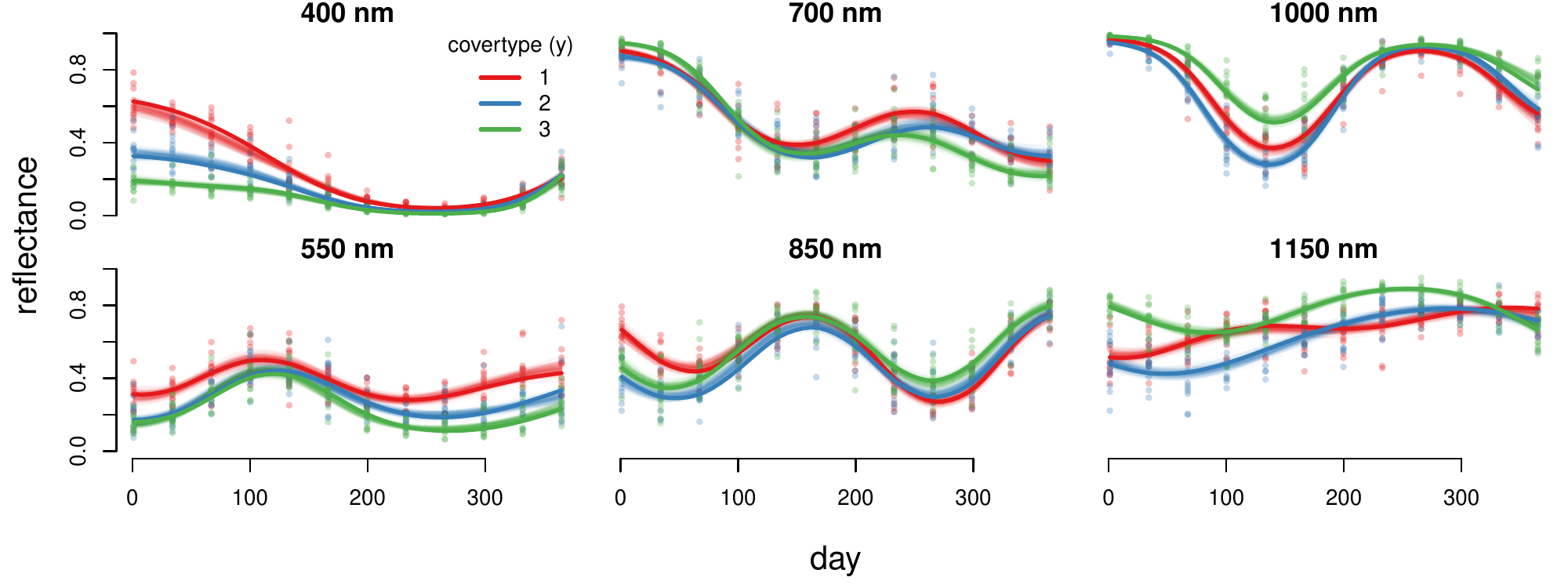}
  \label{fig:reflectance_540}
\end{subfigure}
  \caption{Reflectance surface profiles. Dark, thick curves show values used to simulate the data, and semi-transparent curves show samples from the posterior based on (top row) $J_\mathrm{add} = 20$ and (bottom row) 540. Points show simulated reflectance measurements at sites where covertype was observed ($\boldsymbol{r}_\mathrm{obs}$).}\label{fig:sim_reflectance}
\end{figure}

Supplementary Materials Figure~\ref{fig:sim_reflectance} shows posterior samples of the reflectance profiles in time across all 6 wavelengths for $J_\mathrm{add} = 20$ (top row) and $540$ (bottom row). The two plots are essentially indistinguishable, supporting the second hypothesis in Section~\ref{sec:smoothly_changing} that beliefs about the reflectance surface do not change appreciably with additional reflectance measurements beyond $\boldsymbol{r}_\mathrm{obs}, \boldsymbol{r}_\mathrm{add}$.

Marginal posteriors $\pi(y_j|\boldsymbol{y}_\mathrm{obs}, \boldsymbol{r}_\mathrm{obs}, \boldsymbol{r}_\mathrm{add}, \boldsymbol{r}_j)$ were all highly concentrated near corners of the $K$-simplex, implying very high posterior confidence in one of the $K$ cover types. All posterior modes matched the simulated values of $\boldsymbol{y}_\mathrm{obs}^\mathrm{c}$, suggesting the model could also have very strong predictive power in applications to real data.

\section{Application}\label{sec:application}

\subsection{Data Description}

We applied the proposed model to a data set comprised of $J_\mathrm{obs} = 60$ sites in the Jemez Mountains of New Mexico. The relevant cover types correspond to the three primary types of deciduous vegetation in the region, which are quaking aspen (\textit{Populus tremuloides}), New Mexico locust (\textit{Robinia neomexicana}), and 2 species of oak (\textit{Quercus gambelii}, \textit{undulata}). Subject matter experts selected specific sites and determined vegetation cover type using geo-referenced photographs, which resulted in 21 sites of aspen, 14 of locust, and 25 of oak. Reflectance measurements were collected over all of 2020 by the Sentinel-2A and Sentinel-2B satellites, which together were able to achieve a 2.5 day average revisit interval over the study area. After removing imagery occluded by clouds, the number of observation dates for each site in $\mathcal{J}_\mathrm{obs}$ ranged from 39 to 109, with a median value of 91. Reflectances were measured at 10 wavelengths between 490nm and 2190nm (see Supplementary Materials Table~\ref{tab:bands}). 

Several ($P = 8$) landscape variables of interest were derived for each site based on a 10m resolution digital elevation model provided by the U.S. Geological Survey \citep{USGS_map} including elevation, slope, aspect, ruggedness, and heat load (see Table~\ref{tab:covars}). The goals of the analysis were to estimate the effects of all landscape characteristics on the probability of each cover type, and to use landscape and reflectance information to predict cover type at $\mathcal{J}_\mathrm{obs}^\mathrm{c} = 19,107$ new locations in the region.

\begin{table}[ht]
\caption{Landscape characteristics, $\mathbf{X}$, included in this analysis} \label{tab:covars}
\begin{tabular}{{p{0.25\linewidth} | p{0.7\linewidth}}}
\textbf{Variable} & \textbf{Description} \\ \hline
Elevation (m) & Derived from a digital elevation model with 1 arc second or approximately 30m spatial resolution. \\
Slope (radians) & The rate of change of elevation for each elevation model pixel. Lower values are associated with flatter terrain, higher values with steeper terrain. \\
North exposure & Cosine transformation of aspect, or the compass direction a terrain surface faces.  Values range from -1 (south facing) to 1 (north facing). \\
East exposure & Sine transformation of aspect. Values range from -1 (west facing)  to 1 (east facing). \\
Heat load index (HLI) & Measures potential incident radiation. HLI values range from 0 (coolest, northeast slopes) to 1 (hottest, southwest slopes) \citep{MCCU2002}. \\
Topographic position index (TPI) & Measures the elevation of each pixel relative to its neighborhood. Negative TPI values indicate valley bottoms and positive values indicate ridgetops \citep{DERE2013}. \\
Topographic ruggedness index (TRI) & Measures the amount of elevation difference between adjacent pixels of an elevation model. Low values indicate level topography and high values indicate rugged topography \citep{RILE1999}.
\end{tabular}
\end{table}

Reflectance surfaces were modeled using $L = 70$ thin-plate splines with knots chosen automatically to maintain approximately the same number of observations between each pair of consecutive knots. $M = 35$ thin-plate splines were used to model the spatial random effect, with knots chosen in the same way. As in the simulation study, 1000 iterations we obtained using a slice sampling algorithm after discarding 1000 iterations as burnin. The time required varied from a few hours for $J_\mathrm{add} = 20$ to a few days for $J_\mathrm{add} = 460$ (see Supplementary Materials Figure~\ref{fig:veg_duration}).

\subsection{Results}\label{sec:results}
\begin{figure}[H]
\includegraphics[width=\linewidth]{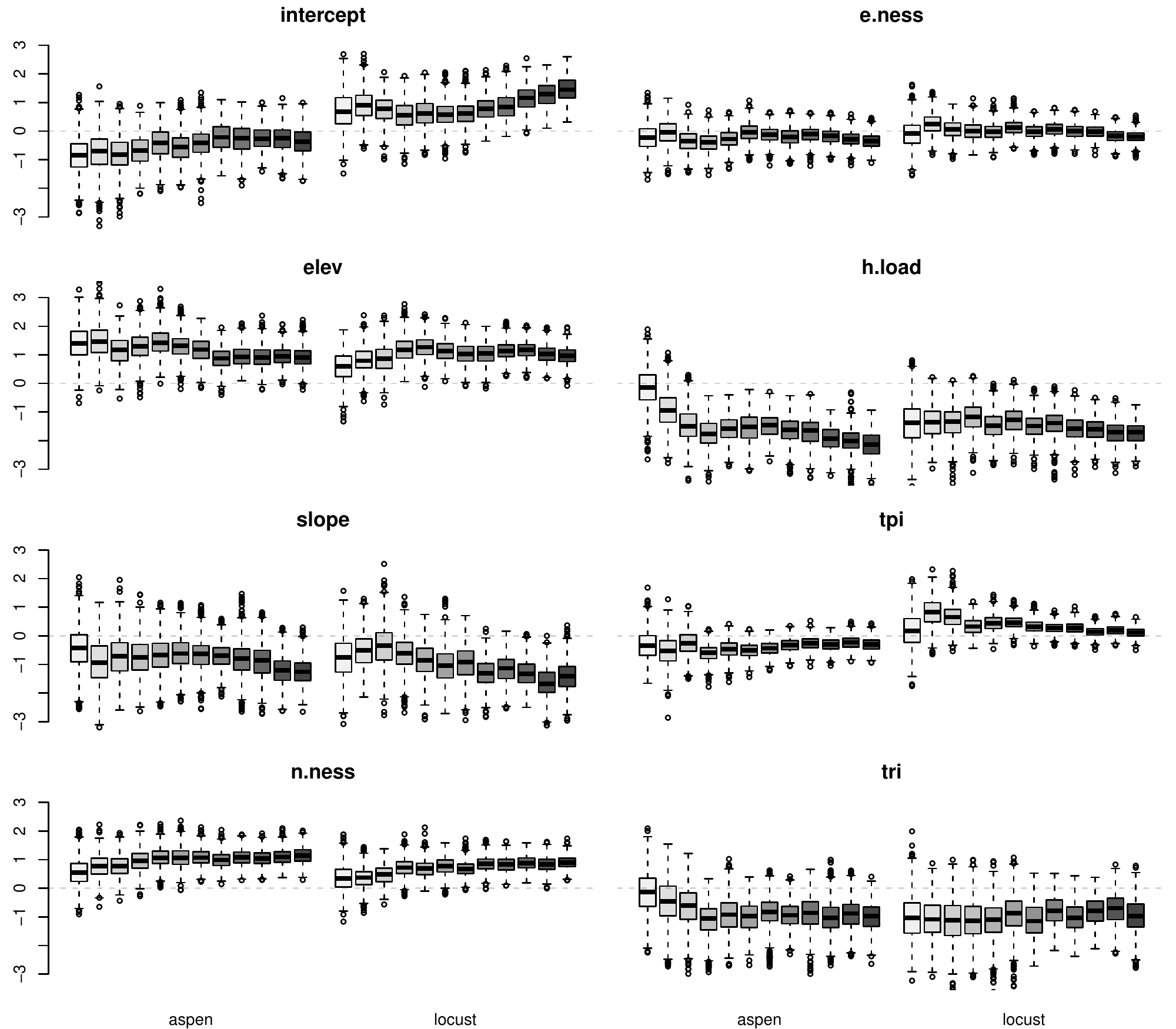}
\caption{Boxplots of marginal posterior distributions for $\mathbf{B}$ across increasing sample sizes. Each subplot corresponds to one of $P$ covariates. Within each subplot, boxplots are grouped by cover type species, with the values for the baseline species, oak, taken to be 0.}
\label{fig:veg_beta}
\end{figure}

Figure~\ref{fig:veg_beta} shows boxplots arranged analogously to Figure~\ref{fig:beta_sim}. Similar to the simulation study, posteriors for $\mathbf{B}$ generally contract as $J_\mathrm{add}$ increases and change smoothly. Most posteriors appear to converge toward stable distributions as $J_\mathrm{add}$ increases, with the possible exceptions of the intercept and heat load. 

\begin{figure}[H]
\includegraphics[width=\linewidth]{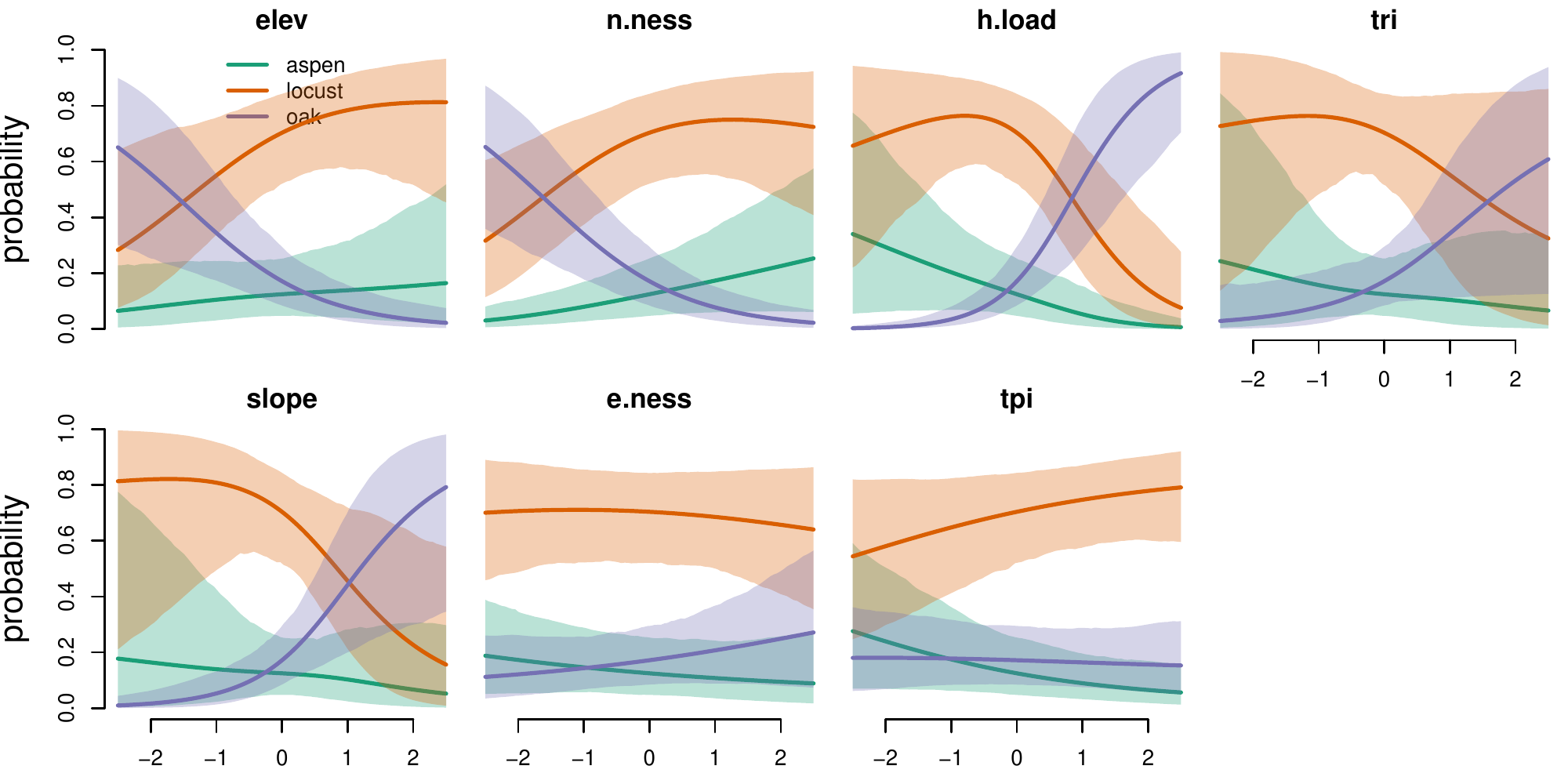}
\caption{Marginal probability curves as functions of each landscape covariate, with all other variables fixed to sample mean values. Lines represent pointwise posterior means, and polygons pointwise equitailed 95\% credible intervals based on 60 labeled sides and 380 additional sites' worth of reflectances.}
\label{fig:veg_prob_curves}
\end{figure}

Figure~\ref{fig:veg_prob_curves} shows marginal probability curves for each landscape variable with other variables assumed fixed at their mean value, and zero spatial random effect. Curves are based on model fits using the maximum $J_\mathrm{add} = 460$ additional sites' worth of data (i.e., the right-most boxplots in Figure~\ref{fig:veg_beta}). A few general patterns emerge: aspen appear most likely to occur at higher elevations on the north side of slopes where the heat loading is lower. Locust appear to be the most abundant type overall, and prefer high elevation, north-facing sites with gentle slopes and low heat loading. Oak species generally fill complementary spaces, which are lower elevation, south-facing, and steeper with higher heat loading.

\begin{figure}[H]
\includegraphics[width=\linewidth]{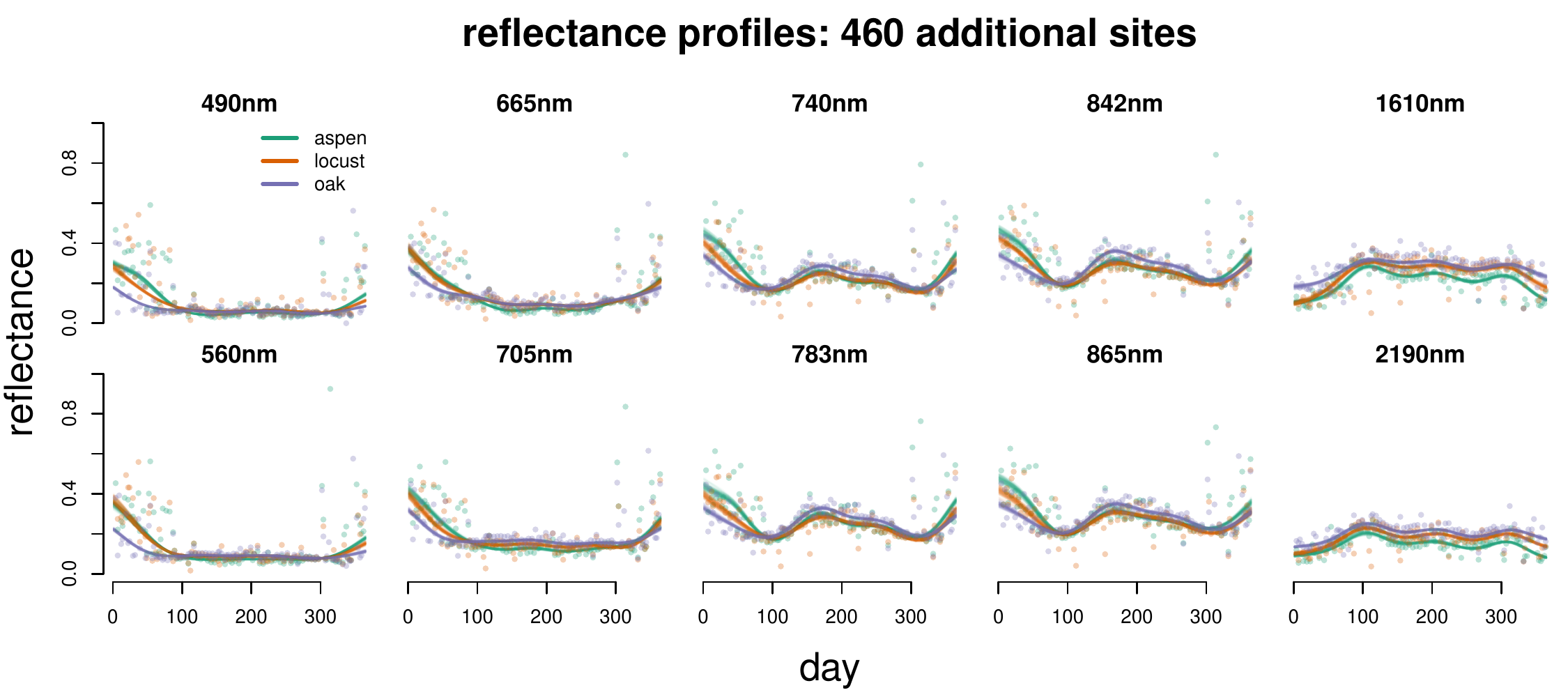}
\caption{Profiles of estimated reflectance surfaces for each species over time, for each observed wavelength level. Lines represent samples from the posterior distribution based on 460 additional sites' worth of reflectances.}
\label{fig:veg_refl_460}
\end{figure}

Figure~\ref{fig:veg_refl_460} shows samples from the marginal posterior distributions of reflectance profiles over time based on the fit using $J_\mathrm{add} = 460$. Strong similarities in the patterns of reflectance among the vegetation types are clear from the high degree of overlap across the functions, but disinctions do exist for certain wavelengths and seasons. Among the three vegetation types, the plots suggest that aspen and locust may be the most difficult pair to distinguish. For example, at wavelengths 740nm and 783nm during the summer (days 172-264), the curve for oak has a small but clear separation from the other two vegetation types.

\begin{figure}[H]
\begin{subfigure}[b]{0.48\textwidth}
  \includegraphics[width=\linewidth]{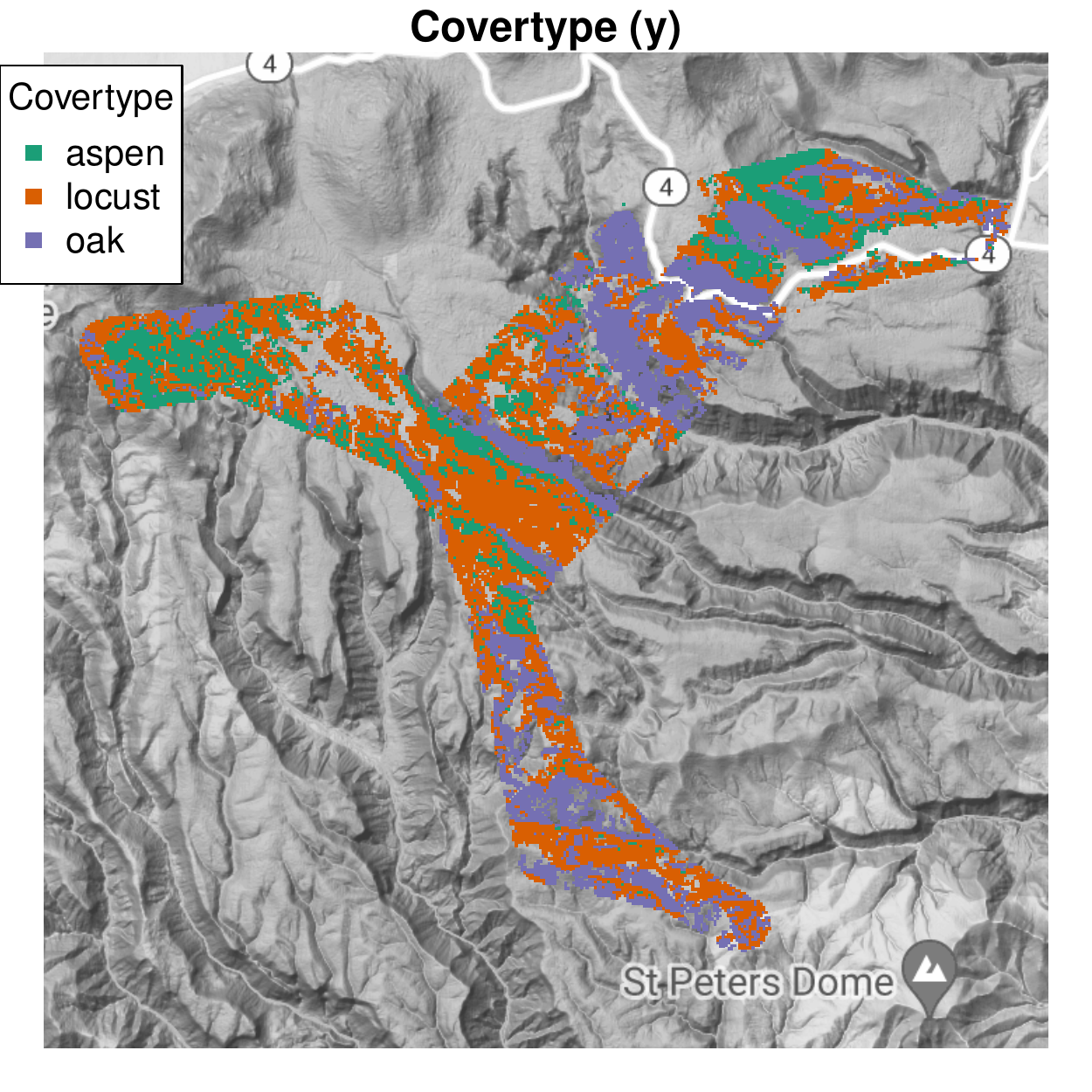}
\end{subfigure}
\begin{subfigure}[b]{0.48\textwidth}
  \includegraphics[width=\linewidth]{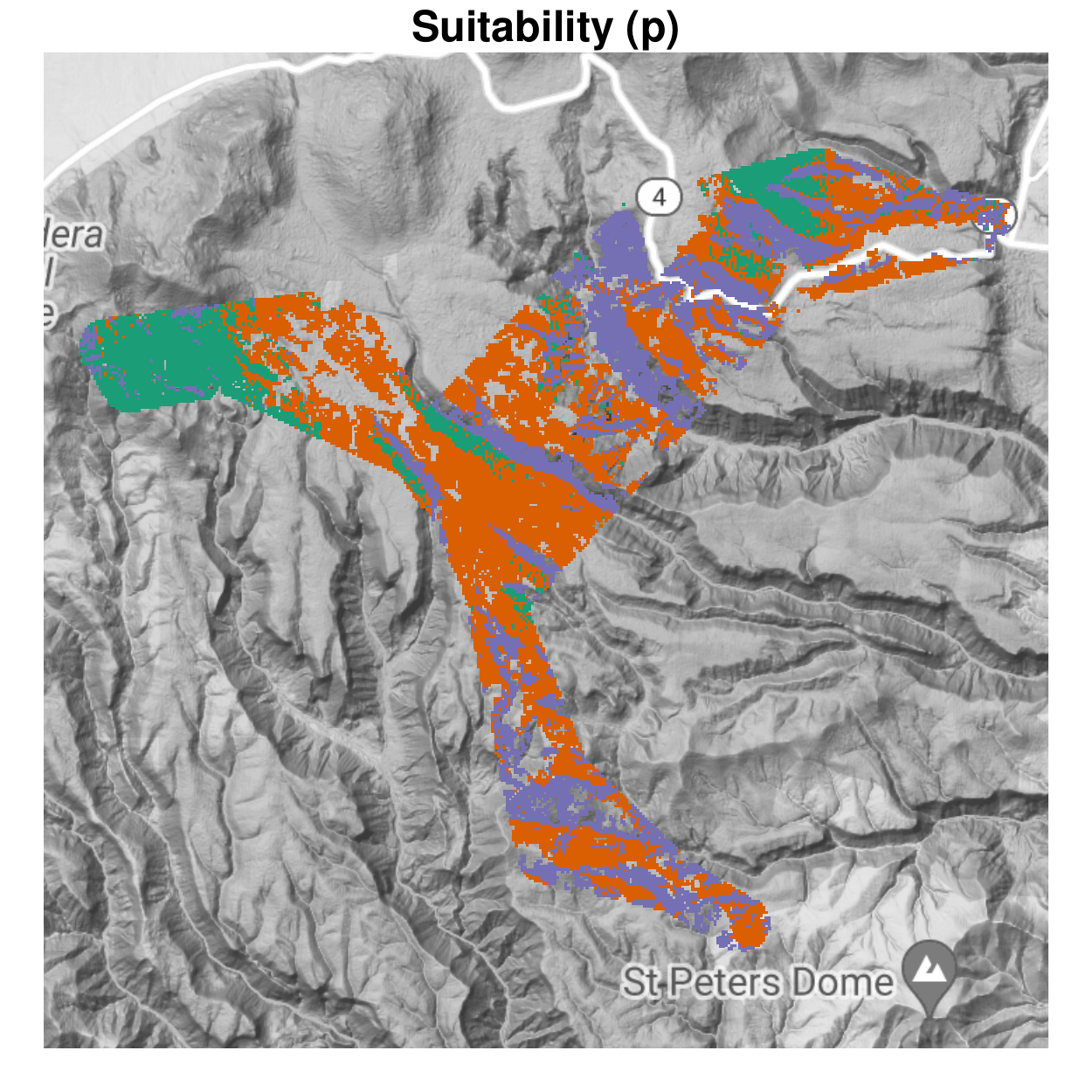}
\end{subfigure}
\caption{Predicted cover type based on posterior of $y_{j}$ (left) and $p_{jk}$ (right).}
\label{fig:veg_pred}
\end{figure}

Figure~\ref{fig:veg_pred} shows a mapped summary of the posterior distribution of $y_{j}$ (left) and $\boldsymbol{p}_{j}$ (right) for all $j \in \mathcal{J}_\mathrm{obs}^\mathrm{c}$ (see Section~\ref{sec:suitability}). Each plot colors sites based on the posterior mode (for $y_j$) or maximum (for $\boldsymbol{p}_{j}$). The left plot was created using Algorithm~\ref{alg:pred_approx}, and represents our belief about the cover type at site $j$ after considering all available data, including the specific reflectances made at site $j$. The driver of this prediction is the collection of reflectance measurements at site $j$, and their compatibility with the estimated reflectance surfaces. The right plot represents the posterior of $\boldsymbol{p}_j$, and can be understood to represent our belief about the cover type at site $j$ using only the measured landscape variables. The compatibility of the observed collection of reflectance measurements at site $j$ and the estimated reflectance surfaces is not used in the creation of the map on the right. 

Across the majority of sites, the left and right maps agree. That is, the type of cover type we would expect to see based on the landscape features match the trend in the measured reflectance at each site. However, there are some locations where the two maps disagree, suggesting regions where the general patterns between cover type and landscape features break down or differ from the region more broadly. For example, the Western-most part of the study area appear to contain more locust-dominated sites and fewer aspen-dominated sites than we would expect based on the local landscape characteristics. Thus, there may be some important unmeasured characteristic about the Western arm of the study region that somehow distinguishes it from the rest of the region.

\section{Discussion}\label{discussion}
We developed and implemented a Bayesian hierarchical model for the analysis of satellite imagery, with the goals of understanding the drivers of species' occurrence distributions and creating maps of predicted cover type for a fixed, known number of species. The proposed model was specified consistent with a causal interpretation in which landscape characteristics give rise to cover type, which in turn give rise to realized measurements of reflectance. Implementing our proposed model with data sets like those available from the European Space Agency's Sentinel-2 mission would not typically be feasible with commonly available computing resources, so we proposed a novel, approximate approach for inference that utilizes as many reflectance observations as possible to understand species' patterns of distribution, and utilized all available reflectance observations when predicting cover type.

Data sets like the one motivating the present work are sometimes called ``tall'' because they include many times more repeated measurements than variables, and many other existing approaches for analyzing them have been proposed \citep[see][for a broad review]{bardenet2017markov}. Our method leverages specific characteristics of our proposed Bayesian hierarchical model in its design to yield reliable predictions of cover type at a modest computational cost, and we do not expect posterior distributions for cover type would change much even if it were possible to forgo the required approximation. In addition, our approach provides inference about the effects of landscape characteristics on cover type probabilities with a clear interpretation as beliefs based on a subset of the total observations. However, we do see room for potential improvement in this second aspect, and many of the successful ideas used with ``tall'' data could inform future research. For instance, multi-stage approaches such as \cite{lunn_fully_2013, hooten_making_2021, hooten_multistage_2023} in which manageable-sized batches of the data are considered in sequence could lead to valid posteriors based on all available data if particle depletion issues can be managed. Alternative directions might include divide-and-conquer type approaches such as \cite{huang_sampling_2005, maclaurin2014firefly} in which data are analyzed in batches and re-combined or coreset methods like \cite{huggins2016coresets} in which weighted subsets represent the full data set.

\section*{Acknowledgments}
We thank Craig Allen, Ellis Margolis and Collin Haffey for discussion on the ecological dlynamics of the post-fire landscape.

\bibliographystyle{apalike}
\bibliography{jemez}

\clearpage
\section*{Supplementary information}
\section{General implementation details}\label{sec:general}
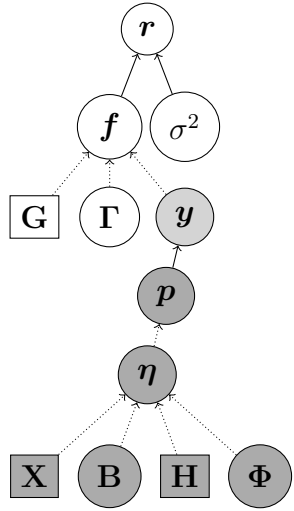
\begin{figure}[H]
\centering
\begin{tikzpicture}
    \node[shape=circle,draw=black] (r) at (2.5,0) {$\boldsymbol{r}$};
    
    \node[shape=circle,draw=black] (f) at (2,-1.3) {$\boldsymbol{f}$};
    \node[shape=circle,draw=black] (sigma) at (3,-1.3) {$\sigma^2$};
  
    \node[shape=rectangle,draw=black] (G) at (1,-2.5) {$\mathbf{G}$};
    \node[shape=circle,draw=black] (gamma) at (2,-2.5) {$\boldsymbol\Gamma$};
    \node[shape=circle,draw=black,fill=black!17] (y) at (3,-2.5) {$\boldsymbol{y}$};

    \node[shape=circle,draw=black,fill=black!33] (prob) at (2.75,-3.55) {$\boldsymbol{p}$};
    
    \node[shape=circle,draw=black,fill=black!33] (eta) at (2.5,-4.6) {$\boldsymbol\eta$};

    \node[shape=rectangle,draw=black,fill=black!33] (x) at (1,-5.95) {$\mathbf{X}$};
    \node[shape=circle,draw=black,fill=black!33] (beta) at (2,-5.95) {$\mathbf{B}$};
    \node[shape=rectangle,draw=black,fill=black!33] (H) at (3,-5.95) {$\mathbf{H}$};
    \node[shape=circle,draw=black,fill=black!33] (psi) at (4,-5.95) {$\boldsymbol\Phi$};

    \path [->] (f) edge node[left] {} (r);

    \path [->, densely dotted] (y) edge node[left] {} (f);
    \path [->, densely dotted] (G) edge node[left] {} (f);
    \path [->, densely dotted] (gamma) edge node[right] {} (f);   
    
    \path [->] (sigma) edge node[right] {} (r);
    
    \path [->] (prob) edge node[right] {} (y);
    
    \path [->, densely dotted] (eta) edge node[right] {} (prob);

    \path [->, densely dotted] (x) edge node[left] {} (eta);
    \path [->, densely dotted] (beta) edge node[right] {} (eta);
    \path [->, densely dotted] (H) edge node[left] {} (eta);
    \path [->, densely dotted] (psi) edge node[right] {} (eta);
\end{tikzpicture}
\caption{Directed acyclic graph representation of hierarchical model. Circular nodes represent stochastic, unknown elements, rectangular ones represent fixed, known elements. Solid lines represent relationships defined by a conditional probability distribution (e.g., $\boldsymbol{r} \sim \pi(\boldsymbol{r} | \boldsymbol{f}, \sigma^2)$), and dotted lines represent deterministic relationships (e.g., $\log(\boldsymbol\eta) = \mathbf{X}\mathbf{B} + \mathbf{H}\mathbf\Phi$). Shadding emphasizes hierarchy in model structure, with covertype acting as a link between landscape characteristics and reflectances.}
\label{fig:DAG}
\end{figure}

\begin{table}[H]
\centering
\caption{Prior distributions for model parameters. Note: $\mathbf{I}_K$ denotes the $K \times K$ identity matrix, and $\mathbf{J}_K$ denotes the $K \times K$ matrix of ones.}
\label{tab:priors}
\begin{tabular}{ll}
\textbf{Parameter} & \textbf{Distribution} \\
\hline
$\boldsymbol\gamma_l, l \in \{1, \dots, L\}$ &
$\mathrm{N}\!\big(\boldsymbol{0}, 0.05\mathbf{\Sigma}_\gamma\big); \; 
  \mathbf{\Sigma}_\gamma = \rho \mathbf{J}_K + (1 - \rho)\mathbf{I}_K$ \\
$\boldsymbol\phi_k, k \in \{1, \dots, K\}$ &
$\mathrm{N}\!\left(\boldsymbol{0}, 0.1\mathbf{I}_M \right)$ \\
$\boldsymbol\beta_k, k \in \{1, \dots, K\}$ &
$\mathrm{N}\!\left(\boldsymbol{0}, \mathbf{I}_P \right)$ \\
$\sigma^2$ & $\mathrm{Gamma}(12, 30)$ \\
\hline
\end{tabular}
\end{table}

\begin{table}[H]
\centering
\caption{Constants}
\label{tab:constants}
\begin{tabular}{lrr}
\textbf{Parameter} & \textbf{Simulate Study} & \textbf{Application} \\
\hline
$J_\mathrm{obs}$ & 36 & 60 \\
$\sum_{j \in \mathcal{J}_\mathrm{obs}} N_j$ & 2,592 & 53,688 \\
$J_\mathrm{add}$ & $\{20, 60, \dots, 540\}$ & $\{20, 60, \dots, 460\}$ \\
$\sum_{j \in \mathcal{J}_\mathrm{obs}^c} N_j$ & 41,472 & 17,428,497 \\
$L$ & 30 & 35 \\
$M$ & 30 & 70 \\
\hline
\end{tabular}
\end{table}

\clearpage 
\section{Simulation study}\label{sec:simulation-study_supp}
\begin{figure}[H]
\centering
\includegraphics[width=\linewidth]{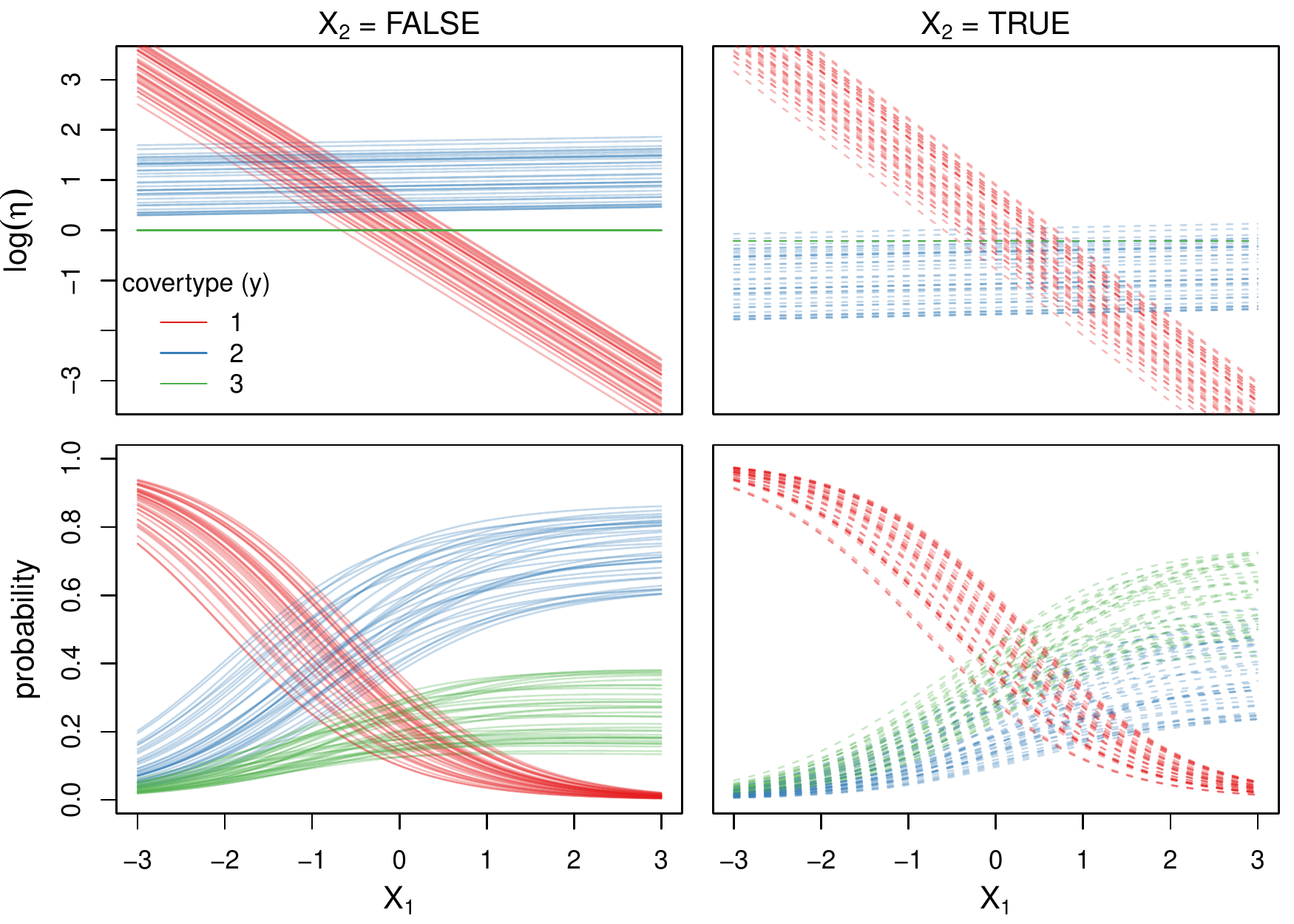}
\caption{Effects of each predictor on covertype (top row) and corresponding probability functions (bottom row), for binary covariate $X_2$ taking on values of FALSE/0 (left column) and TRUE/1 (right column).}
\label{fig:sim_prob_known}
\end{figure}

\begin{sidewaysfigure}[ht]
\includegraphics[width=\linewidth]{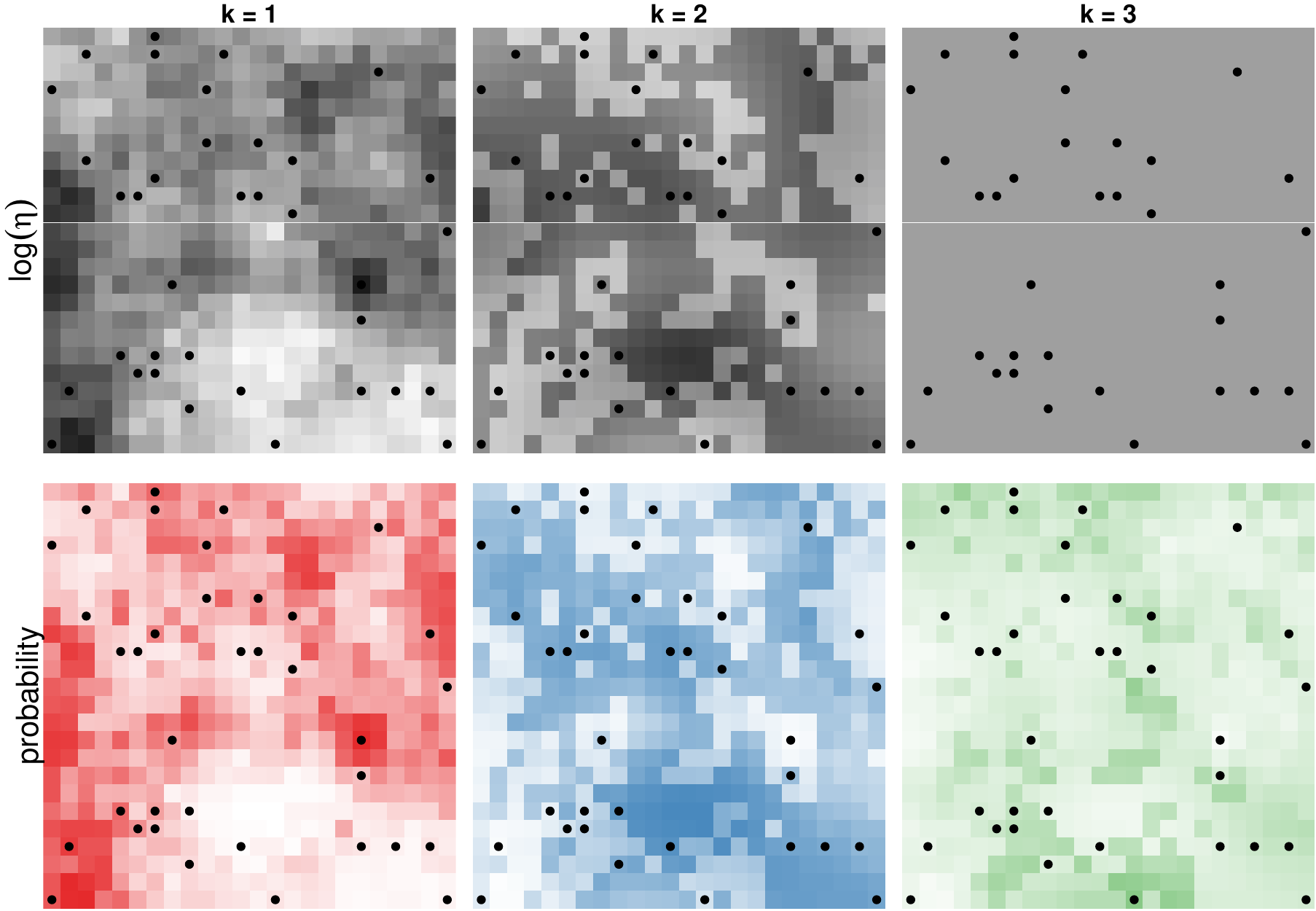}
\caption{Combined log-linear covariate and spatial random effects at each site (top row) and corresponding probabilities (bottom row). Darker colors correspond to greater values. Black points represent locations where covertype was measured and used in model fitting.}
\label{fig:sim_spatial_probs}
\end{sidewaysfigure}

\begin{figure}[H]
\centering
\includegraphics[width=0.5\linewidth]{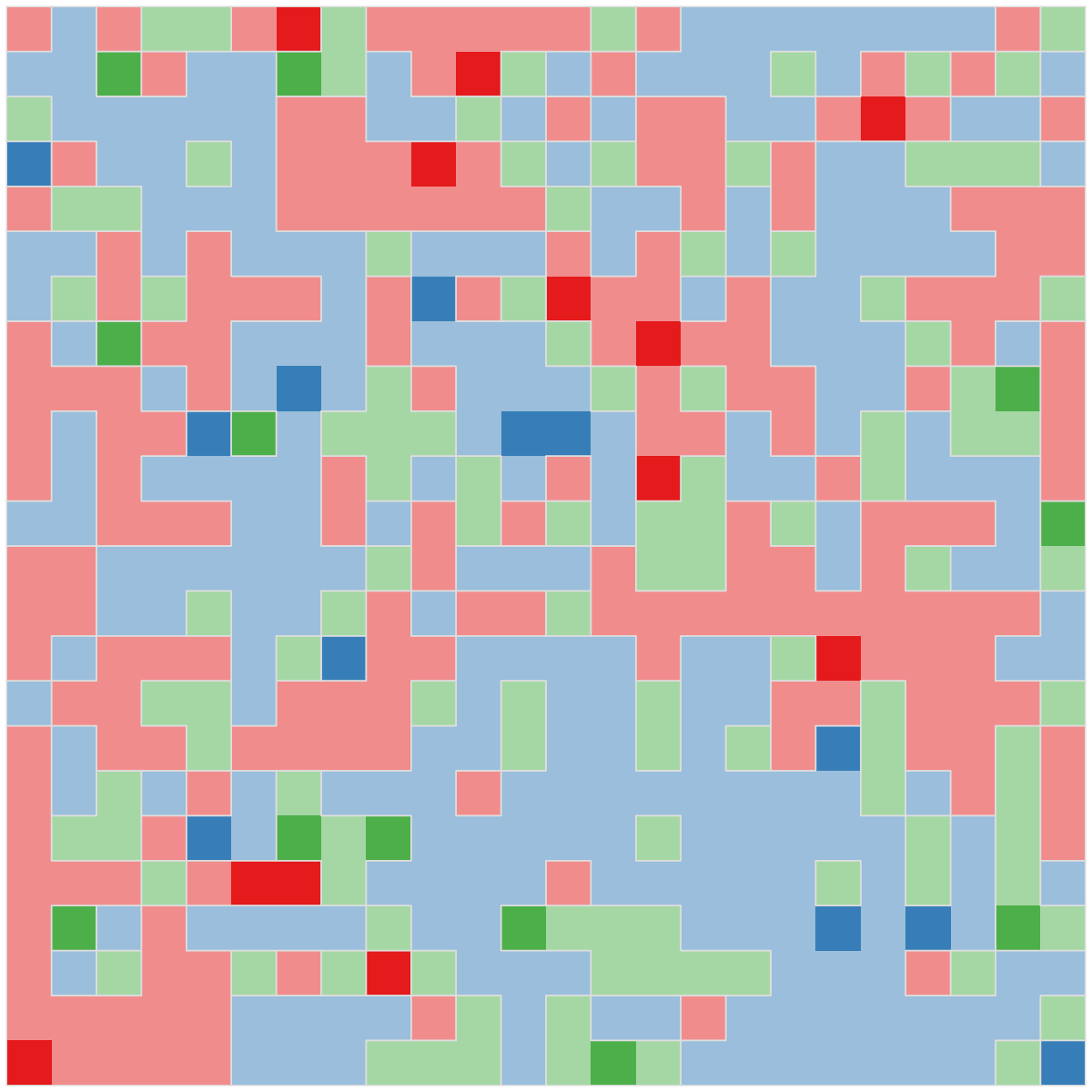}
\caption{Simulated covertype for each location in the simulation study. Dark pixels represent locations where covertype was measured and used in model fitting.}
\label{fig:sim_cover}
\end{figure}

Supplementary Materials Figure~\ref{fig:sim_prob_known} shows the linear effects $\boldsymbol{x}_j^\prime \boldsymbol\beta_k$ (top row) and resulting marginal probabilities (bottom row) of a site having a given cover type as functions of the two synthetic covariates. Each curve represents the probability of observing a particular cover type for the 36 sites in $\mathcal{J}_\mathrm{obs}$. The variation among curves depicts the spatial random effect given by $\boldsymbol{h}_j^\prime\boldsymbol\phi_k$. Note that the linear and spatial effects for cover type $k = 3$ are fixed to be zero to ensure identifiability (top row), yet the resulting probability curves vary for all cover types (bottom row). Supplementary Materials Figure~\ref{fig:sim_spatial_probs} depicts the combined log-linear covariate and spatial effects at each site (top row) as well as the resulting marginal cover type probabilities (bottom row). Of the 576 total locations, $J_\mathrm{obs} = 36$ were chosen completely at random to represent locations where cover type was observed, under the constraint that $\boldsymbol{y}_\mathrm{obs}$ included 12 cases of each cover type. Supplementary Materials Figure~\ref{fig:sim_cover} shows the realized pattern of cover types with darker pixels corresponding to those randomly selected to represent sites with recorded cover type. 

\subsection{Covertype}

\clearpage 

\subsection{Reflectance}

\subsection{Estimated marginal probability curves}
\begin{figure}[H]
\includegraphics[width=\linewidth]{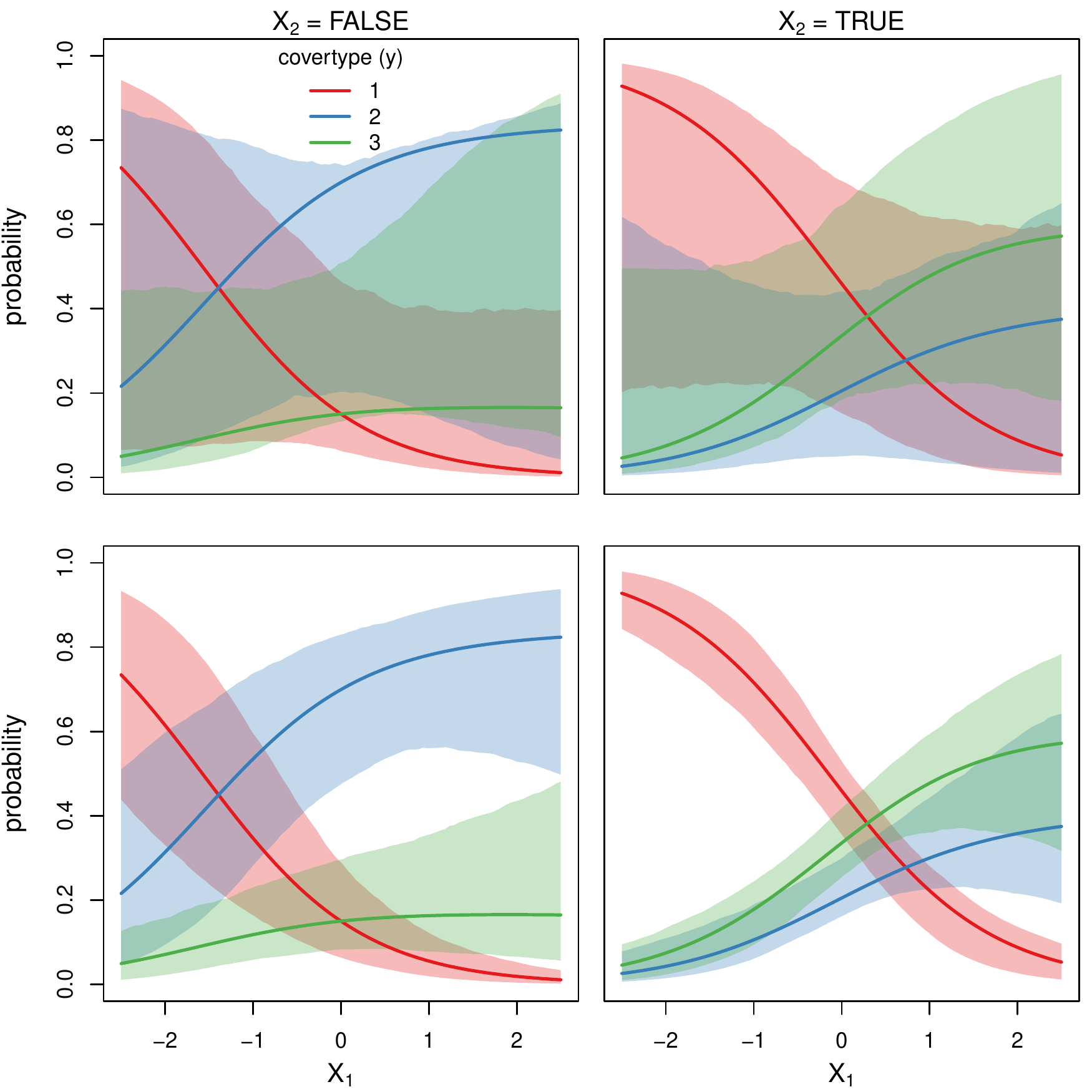}
\caption{Estimated probability functions based on the mininum ($J_\mathrm{add} = 20$) (top row) and maximum ($J_\mathrm{add} = J_\mathrm{obs}^\mathrm{c} = 540$) number of observed reflectances (bottom row). Solid lines represent the probabilities for each covertype ignoring the spatial random effects used to generate the simulated data, and polygons represent pointwise 95\% credible intervals for these probabilities.}
\label{fig:sim_prob}
\end{figure}

\clearpage
\section{Application}\label{sec:application_supp}
\begin{table}[H]
\caption{Sentinel-2 bands included in this analysis} \label{tab:bands}
\begin{tabular}{lllll}
\textbf{Band} & \multicolumn{1}{l|}{\textbf{Name}}         & \textbf{\begin{tabular}[c]{@{}l@{}}Central \\ Wavelength (nm)\end{tabular}} & \textbf{Bandwidth (nm)} & \textbf{\begin{tabular}[c]{@{}l@{}}Ground \\ Resolution (m)\end{tabular}} \\
\multicolumn{5}{c}{\cellcolor[HTML]{C0C0C0}{\color[HTML]{333333}\textbf{Sentinel-2 Bands Included In Analysis}}} \\
2 & \multicolumn{1}{l|}{Blue} & 490 & 65 & 10 \\
3 & \multicolumn{1}{l|}{Green} & 560 & 35 & 10 \\
4 & \multicolumn{1}{l|}{Red}  & 665 & 30 & 10 \\
5 & \multicolumn{1}{l|}{Red Edge 1} & 705 & 15 & 20 \\
6 & \multicolumn{1}{l|}{Red Edge 2} & 740 & 15 & 20 \\
7 & \multicolumn{1}{l|}{Red Edge 3} & 783 & 20 & 20 \\
8 & \multicolumn{1}{l|}{Near-Infrared} & 842 & 115 & 10 \\
8A & \multicolumn{1}{l|}{Narrow Near-Infrared}  & 865 & 20 & 20 \\
11 & \multicolumn{1}{l|}{SWIR-2} & 1610 & 90 & 20 \\
12 & \multicolumn{1}{l|}{SWIR-3} & 2190 & 180 & 20 \\
\multicolumn{5}{c}{
\cellcolor[HTML]{C0C0C0}\textbf{Sentinel-2 Bands Excluded From Analysis}} \\
1 & \multicolumn{1}{l|}{Coastal aerosol} & 443 & 20 & 60 \\
9 & \multicolumn{1}{l|}{Water Vapor} & 940 & 20 & 60 \\
10 & \multicolumn{1}{l|}{SWIR - Cirrus} & 1375 & 20 & 60                
\end{tabular}%
\end{table}

\begin{figure}[H]
  \begin{subfigure}[b]{0.48\textwidth}
  \includegraphics[width=\textwidth]{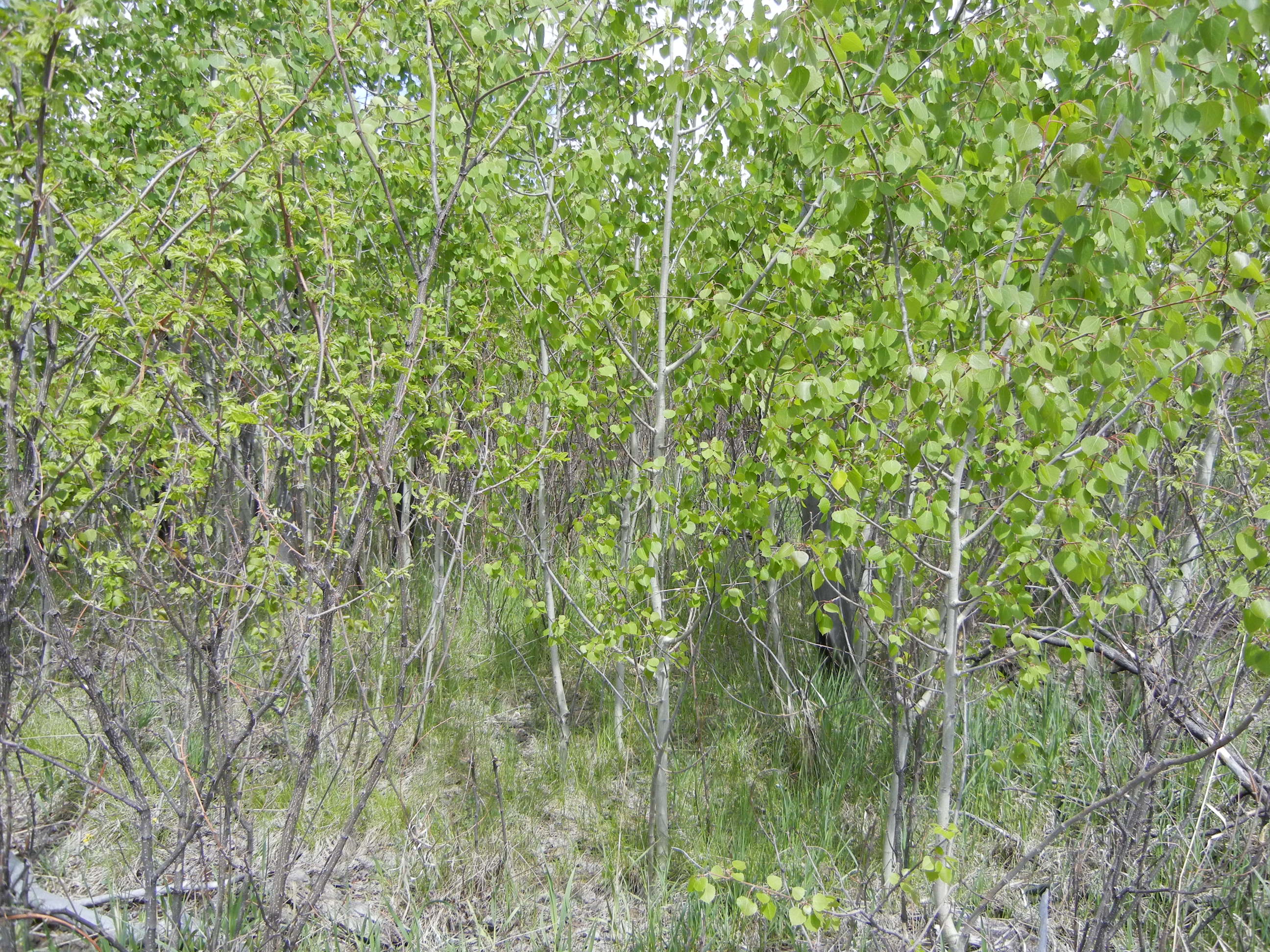}
  \caption{quaking aspen}
  \end{subfigure}
  \begin{subfigure}[b]{0.48\textwidth}
  \includegraphics[width=\textwidth]{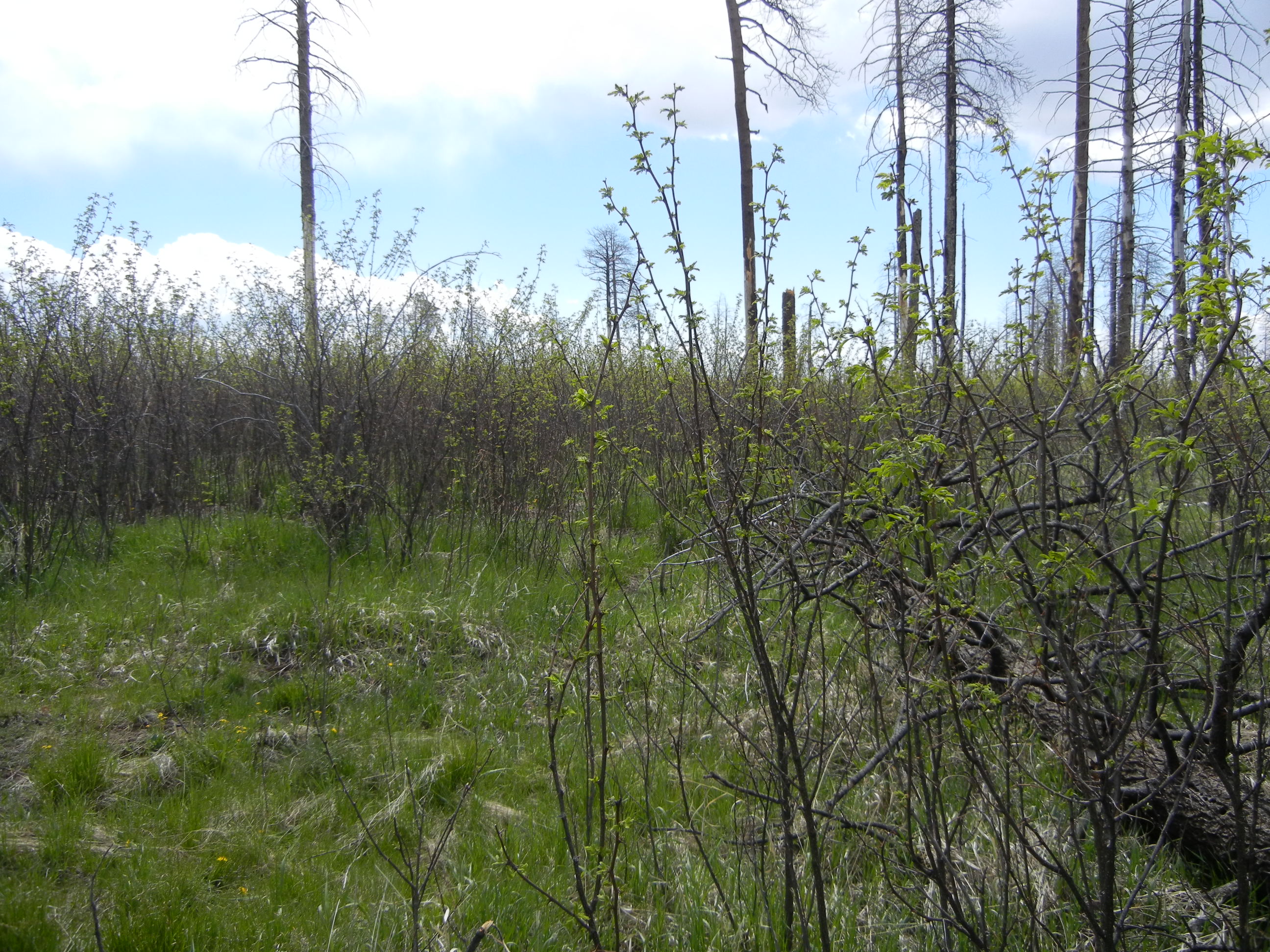}
  \caption{New Mexico locust}
  \end{subfigure}
  \begin{subfigure}[b]{0.68\textwidth}
  \includegraphics[width=\textwidth]{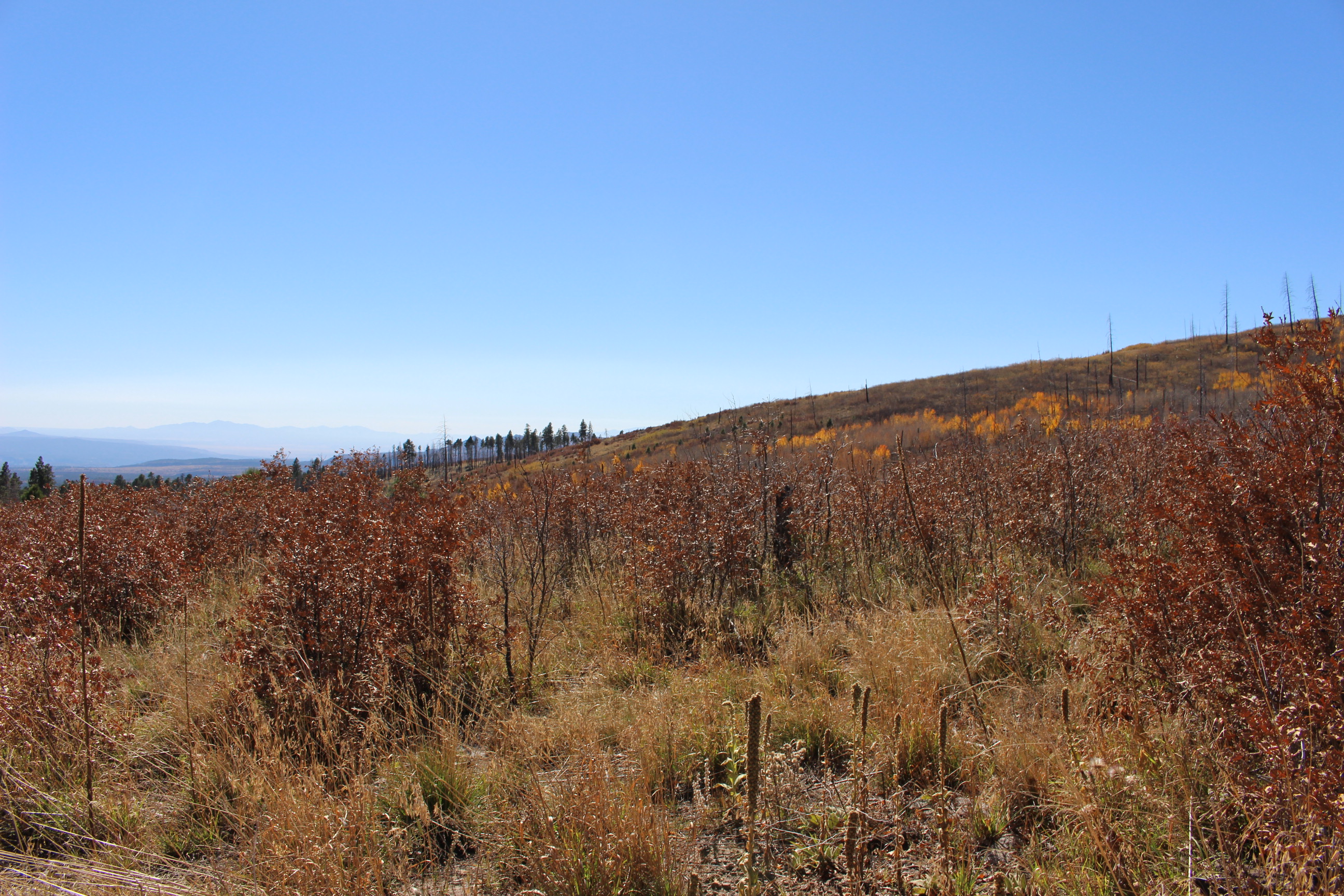}
  \caption{oak}
  \end{subfigure}
  \caption{Example georeferenced photographs collected in the field used for identifying covertype for $j \in \mathcal{J}_\mathrm{obs}$.}
\label{fig:study_photos}
\end{figure}

The majority of reference sites used in our analysis are of low-density young stands (see, Figure~\ref{fig:study_photos}). Considering that a remotely sensed pixel has a surface footprint between $10 \times 10$ and  $60 \times 60$ meters depending on instrument, these photographs contextualize the challenge of classifying low density stands. They also are representative of the heterogeneous species distribution patterns characterizing the study area landscape overall. 

\begin{figure}[H]
\centering
\includegraphics[width = 0.7\linewidth]{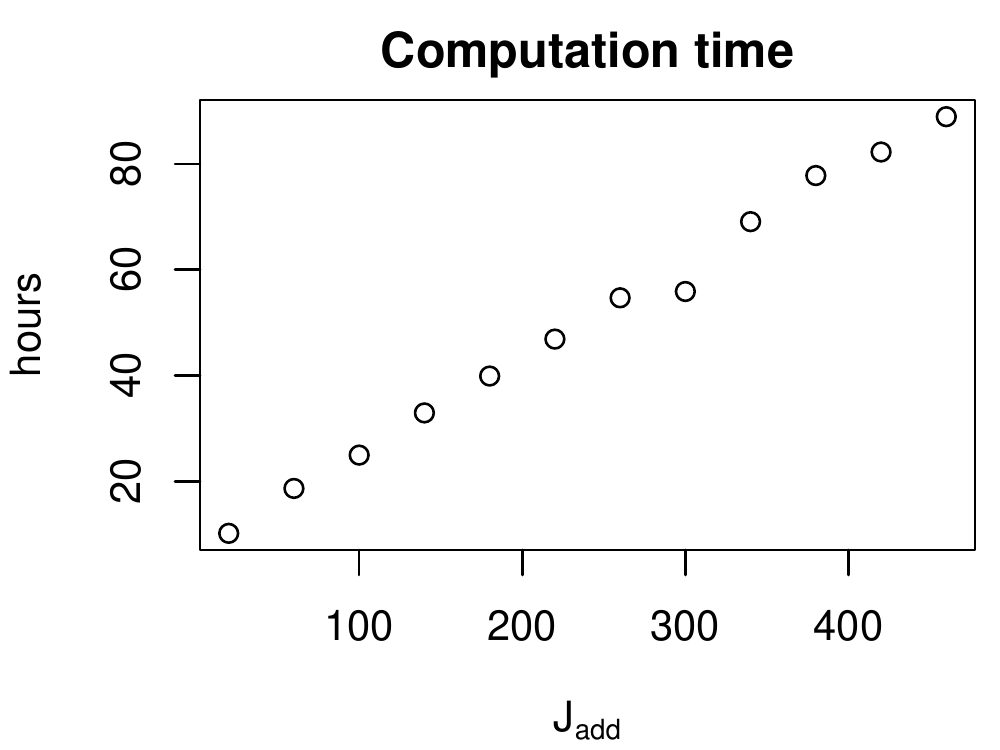}
\caption{Time required to obtain 1000 samples from posterior distribution including burnin.}
\label{fig:veg_duration}
\end{figure}

\begin{figure}[H]
\includegraphics[width=\linewidth]{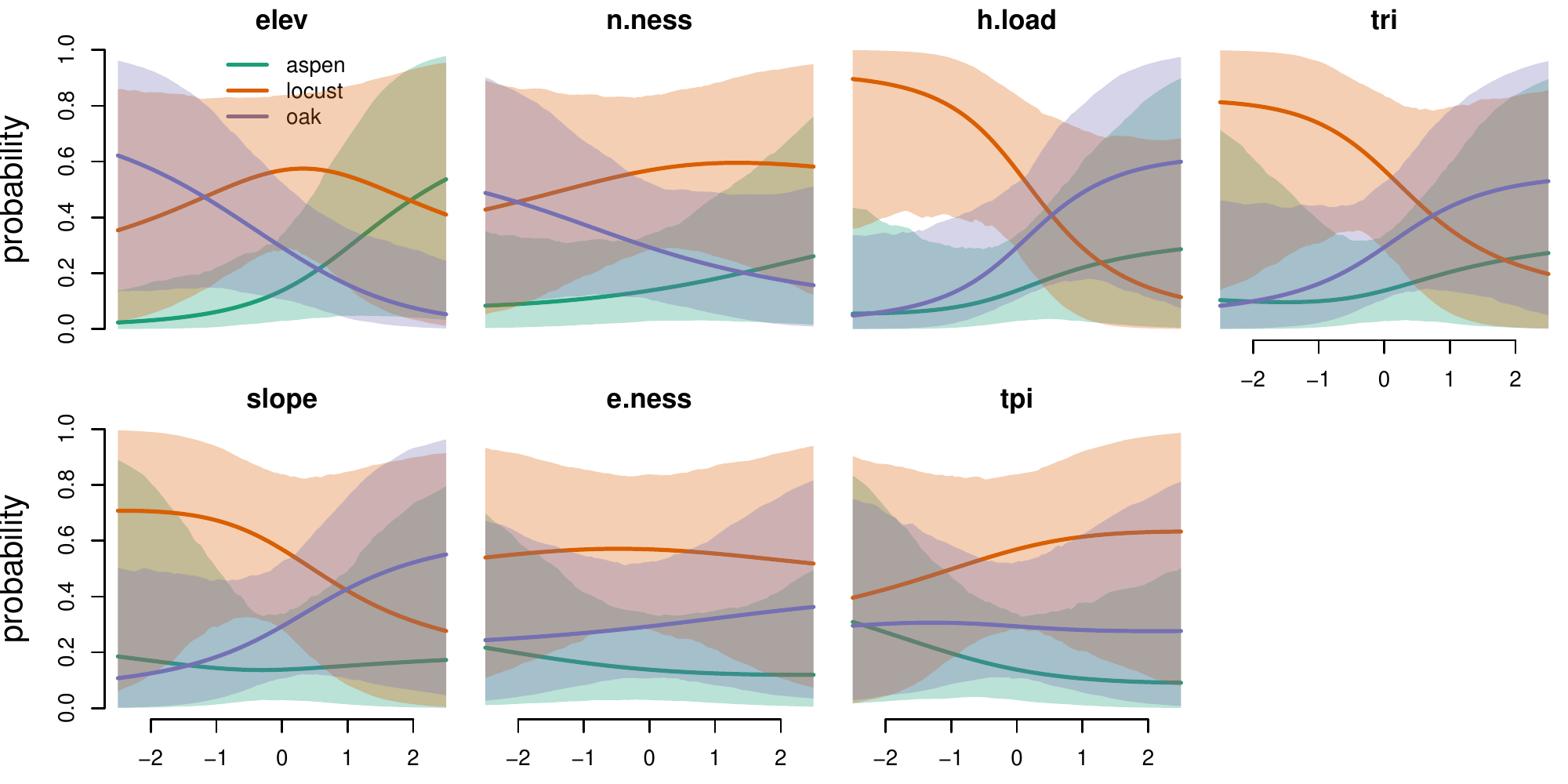}
\caption{Marginal probability curves as functions of each landscape covariate, with all other variables fixed to sample mean values. Lines represent pointwise posterior means, and polygons pointwise equitailed 95\% credible intervals based on 60 labeled sides and 20 additional sites' worth of reflectances.}
\label{fig:veg_prob_20}
\end{figure}

\begin{figure}[H]
\includegraphics[width=\linewidth]{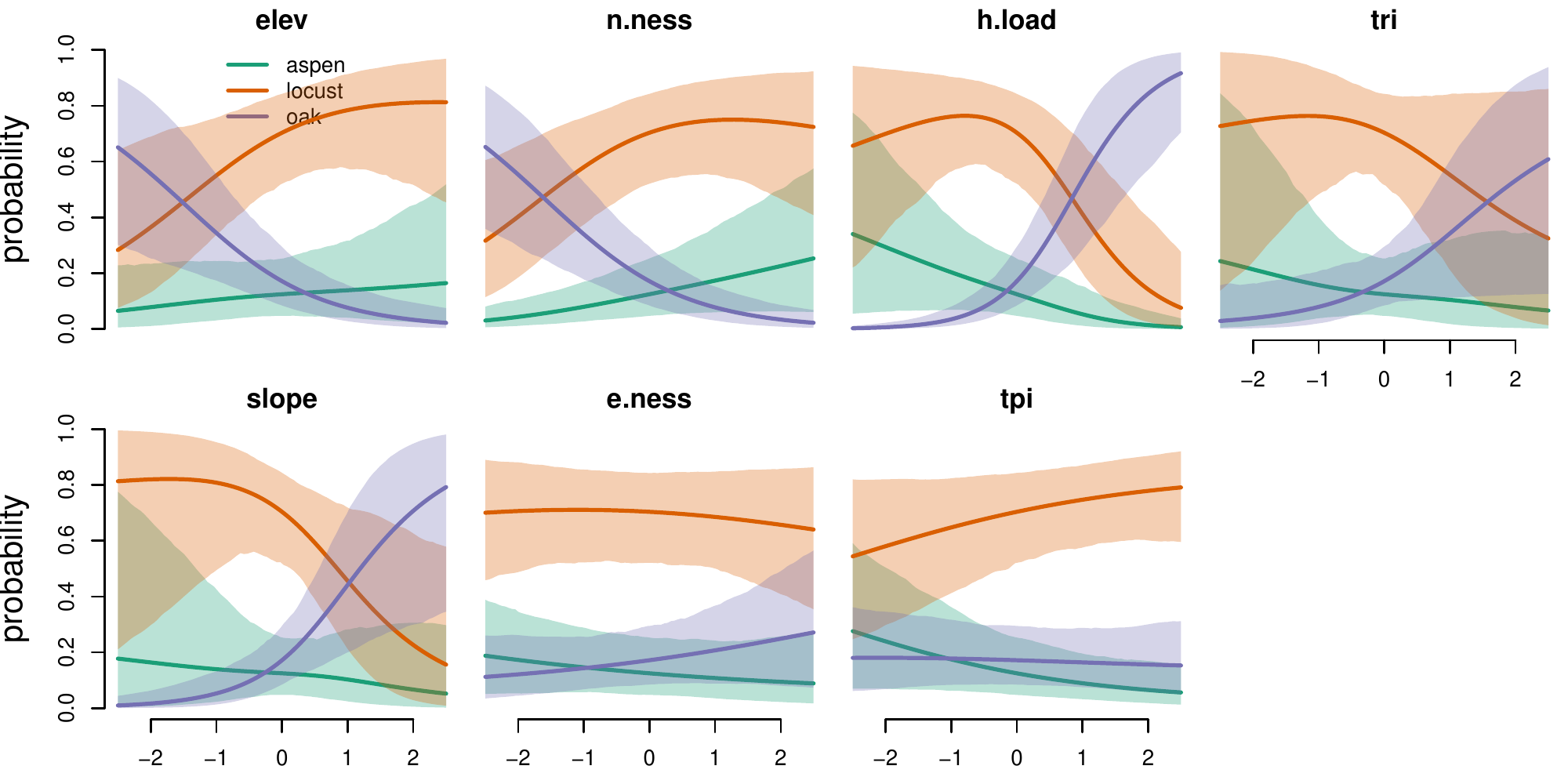}
\caption{Marginal probability curves as functions of each landscape covariate, with all other variables fixed to sample mean values. Lines represent pointwise posterior means, and polygons pointwise equitailed 95\% credible intervals based on 60 labeled sides and 380 additional sites' worth of reflectances.}
\label{fig:veg_prob_460}
\end{figure}

\begin{sidewaysfigure}[ht]
\includegraphics[width=\linewidth]{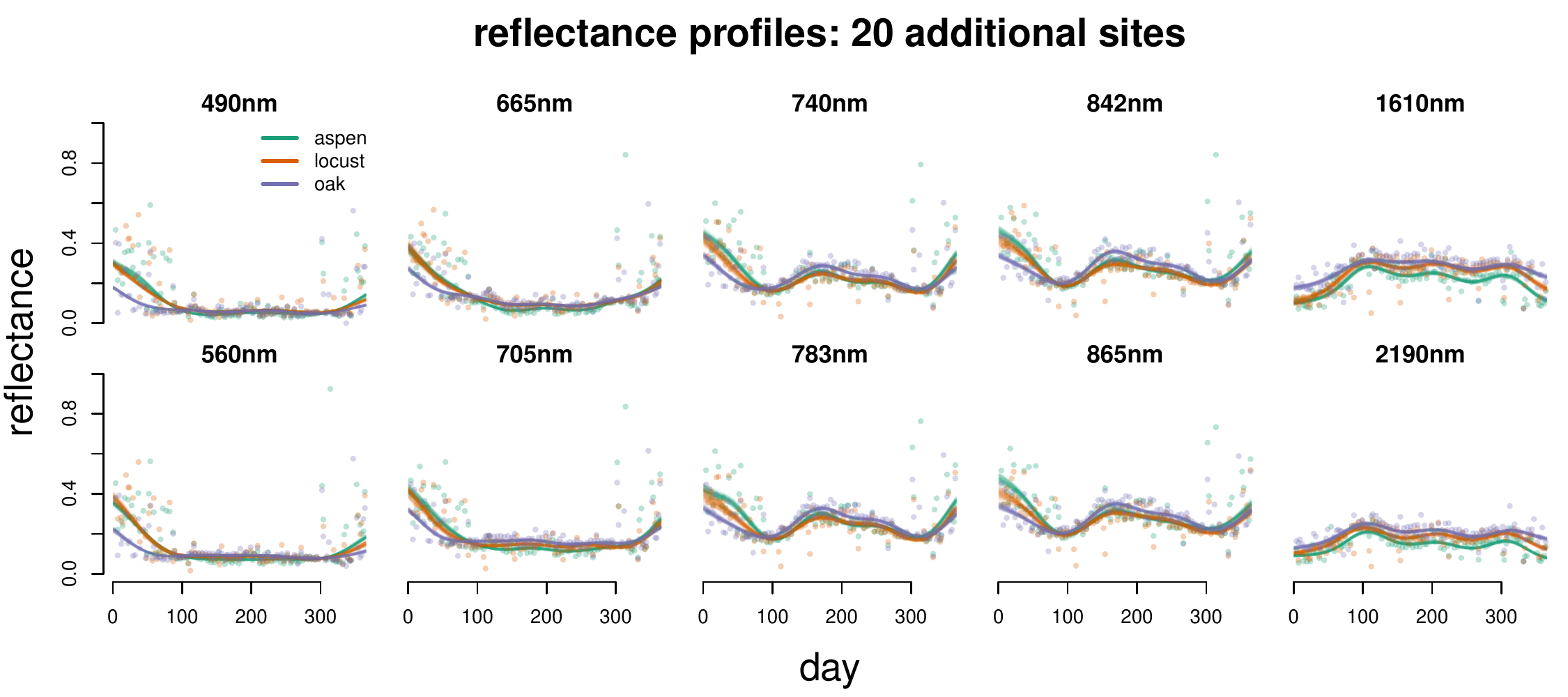}
\caption{Profiles of estimated reflectance surfaces for each species over time, for each observed wavelength level. Lines represent samples from the posterior distribution based on 20 additional sites' worth of reflectances.}
\label{fig:veg_refl_20}
\end{sidewaysfigure}

\end{document}